\documentclass{JHEP}
\pdfoutput=1
\usepackage{amsmath}
\usepackage{graphicx}
\usepackage{multirow}

\usepackage{amsmath,amstext,amsfonts,amsbsy,amssymb,amscd,bbm,epsfig,lscape,cite}

\newcommand{\be}{\begin{equation}}
\newcommand{\ee}{\end{equation}}
\newcommand{\bea}{\begin{eqnarray}}
\newcommand{\eea}{\end{eqnarray}}

\newcommand{\He}{$^6$He~}
\newcommand{\Ne}{$^{18}$Ne~}

\title{Precision on leptonic mixing parameters at future neutrino oscillation experiments}
\preprint{
  CERN-PH-TH/2012-071\\
	IFIC/12-21\\
	IFT-UAM/CSIC-12-24\\
  EURONU-WP6-12-47\\
	}

\author{P.~Coloma$^1$, A.~Donini$^{2,3}$, E. Fern\'andez-Mart\'inez$^4$,  P.~Hern\'andez$^3$\\
(1) Center for Neutrino Physics, Department of Physics, Virginia Tech, Blacksburg, VA 24061, USA \\
(2) Instituto de F\'{\i}sica Te\'orica UAM/CSIC, \\
      Calle Nicol\'as Cabrera 13-15, E-28049 Madrid, Spain \\
(3)  Instituto de F\'{\i}sica Corpuscular, CSIC-Universitat de Val\`encia,\\
       Apartado de Correos 22085, E-46071 Valencia, Spain\\	
(4) Theory Division, CERN, 1211 Gen\`eve 23, Switzerland\\ 
}

\abstract{
We perform a comparison of the different  future neutrino oscillation experiments based on the achievable precision 
in the determination of the fundamental parameters $\theta_{13}$ and the CP phase, $\delta$, assuming that $\theta_{13}$ is in the range
indicated by the recent Daya Bay measurement. We study the non-trivial dependence of the error on $\delta$  on its true value. When matter effects are small, the largest error is found at the points where CP violation is maximal, and the smallest at the CP conserving points. The situation is different when matter effects are sizable. As a result of this effect,  the comparison of the physics reach of different experiments on the basis
of the CP discovery potential, as usually done, can be misleading.  
We have compared various proposed super-beam, beta-beam and neutrino factory setups on the basis of the relative precision of $\theta_{13}$ and the error on $\delta$. Neutrino factories, both high-energy or low-energy, outperform alternative beam technologies. 
An ultimate precision on $\theta_{13}$ below 3\% and an error on $\delta$ of $\leq 7^\circ$ at $1\sigma$ (1 d.o.f.) can be obtained at a neutrino factory.
}

\begin{document}

\section{Introduction}
\label{sec:intro}

The first results of Daya Bay \cite{An:2012eh}  provide the first measurement of the angle $\theta_{13}$. The T2K experiment had earlier published a $\sim 2.5 \sigma$ hint of a non-vanishing angle~\cite{Abe:2011sj}, also confirmed at a lesser statistical significance  
 by the first results of Double Chooz \cite{Abe:2011fz} and by the 
$\nu_e$ appearance measurement of MINOS \cite{Adamson:2011qu}. Previous analyses had already hinted that $\theta_{13} \neq 0$ could improve the $\chi^2$ of global fits, in particular the agreement between solar and KamLAND data~\cite{GonzalezGarcia:2010er,Fogli:2011qn}. 

The angle $\theta_{13}$ is a fundamental parameter of the Standard Model. As such, we would like to measure it with as good precision as  possible, and hopefully with the same precision as its equivalent mixing angle in the quark sector. In fact, given the large hierarchy between neutrino masses and the remaining fermion masses, it is of the utmost importance to test the lepton flavour sector of the Standard Model, since it could unveil the mechanism of neutrino mass generation and the explanation of this hierarchy.  Furthermore, $\theta_{13}$ is also the missing link to a new source of CP violation in the Standard Model. Leptonic CP violation could have profound consequences in particle physics and cosmology, as it could be related to the origin of the matter-antimatter asymmetry of the Universe \cite{Fukugita:1986hr}.

In the last ten years, many different strategies have been put forward to measure $\theta_{13}$ and to discover leptonic CP violation in future experiments \cite{Bandyopadhyay:2007kx}. Improving the statistics and reducing the background systematic errors of conventional neutrino beams would be mandatory had $\theta_{13}$ turned out to be very small ($\theta_{13} \lesssim 3^\circ$).  These can be achieved with purer neutrino beams, such as those that could be produced in a neutrino factory~\cite{Geer:1997iz,DeRujula:1998hd,Apollonio:2002en,GomezCadenas:2002fz} ({\em i.e.} a muon storage ring) or in a beta-beam~\cite{Zucchelli:2002sa} ({\em i.e.} radioactive ion storage ring). However, for values of $\theta_{13}$ as large as recent data indicates ($ \theta_{13} \sim 9^\circ$) more intense conventional neutrino beams may also have a good chance to perform these measurements. It is therefore important to compare how these very different approaches will perform in the task. 

In most previous studies, it has been common to compare the performance in terms of the discovery potential for a non-vanishing $\theta_{13}$ or for CP violation, {\em i.e.} depicting the areas of the parameter space in $(\theta_{13}, \delta)$ where $\theta_{13}$ could be distinguished from zero or $\delta$ from CP-conserving values ($0, \pi$) at a given confidence level. In such comparisons, facilities with more intense and purer beams outperform the others very significantly~\cite{NF:2011aa}. On the other hand, for a largish $\theta_{13}$ it makes more sense to perform the comparison in terms of the precision achievable on those parameters, 
since the discovery of the unknown parameters is almost granted. This is the goal of the present paper.

We have considered most of the  setups previously discussed in the literature, classifying them according to three types of neutrino beams: conventional or super-beams, 
beta-beams and neutrino factories. Among each class we compare different experiments that might involve different average neutrino energies, different baselines and/or different detector technologies. 
 The comparison will be based on two quantities: the relative error on the angle $\theta_{13}$ and the absolute error on the CP phase $\delta$. We will show our results for values of $\theta_{13}$ inside the $3 \sigma$ region preferred by the recent Daya Bay results. We will assume that, either each experiment will be able to distinguish the neutrino mass hierarchy by itself, or that it can do it in combination with future atmospheric neutrino measurements (particularly at its own detector~\cite{Huber:2005ep,Campagne:2006yx}) and/or with the present generation of neutrino oscillation experiments~\cite{TabarellideFatis:2002ni,Bernabeu:2003yp,PalomaresRuiz:2004tk,Indumathi:2004kd,Petcov:2005rv,Arumugam:2005nt,Samanta:2006sj,Gandhi:2007td,Mena:2008rh,Gandhi:2008zs,Samanta:2009qw,Blennow:2012gj} 
(T2K, MINOS, NOVA, INO, reactors).

The issue of precision on leptonic mixing parameters has been addressed in earlier analyses of the performance of super-beams and neutrino factories. In particular, curves of the error on $\delta$ as a function of $\delta$, for fixed $\theta_{13}$, were first shown in Refs.~\cite{ Winter:2003ye,Huber:2004gg}). 
 In this paper, we extend those studies in various ways. We identify the main features that explain the striking dependence of $\Delta\theta_{13}$ and $\Delta\delta$ 
 on the true values of the parameters, focusing on the parameter range implied by Daya Bay result. We have also widened and updated the range of experiments 
 considered and performed a systematic comparison of their physics performance on the basis of precision.

The structure of the paper is as follows: in Sec. \ref{sec:setups} we introduce our precision observables and briefly summarize the facilities that will enter in the comparison; 
Sec. \ref{sec:precision} contains a discussion on the dependence of the precision observables on the true values of $\theta_{13}$ and $\delta$ in their presently allowed ranges; 
the numerical results of the comparison of the different setups are summarized  in Sec. \ref{sec:results}; we eventually conclude in Sec. \ref{sec:concl}.

\section{Observables and setups}
\label{sec:setups}

\subsection{Precision observables and simulation details}

The goal of this paper is the study of the performances of several facilities in terms of the attainable precision 
in the determination of  the parameters $\theta_{13}$ and $\delta$. The two observables that we have considered are
the relative error on $\theta_{13}$, $\Delta_r\theta_{13} \equiv \Delta\theta_{13} / \theta_{13}$, and the absolute error on $\delta$, $\Delta\delta$.

For a fixed value of the true parameters $\theta_{13}$ and $\delta$, the absolute errors
$\Delta\theta_{13}$ and $\Delta\delta$ are defined as one half of the reconstructed 1$\sigma$ range (1 d.o.f.) for the corresponding variable, 
after marginalizing over all other oscillation and nuisance parameters\footnote{Note that, only if the confidence region is symmetric around the best fit, do the upper and lower error bars coincide with $\Delta X$. In any case, $2 \Delta X$ always corresponds to the sum of the upper and lower error bars. Furthermore, for strongly non-gaussian situations, such as the presence of degeneracies disfavoured only at the $1 \sigma$ level, higher confidence level regions may significantly differ from a naive rescaling of $\Delta X$.}.
The $\chi^2$ has been computed using the GLoBES 3.0 software \cite{Huber:2004ka,Huber:2007ji}.

As we will see, the precision on $\theta_{13}$ and $\delta$ depends rather significantly on the true values of the parameters $ \theta_{13}$ and, especially, $ \delta$.
For this reason, $\Delta_r \theta_{13}$ and $\Delta \delta$ are shown as functions of $\theta_{13}$ and $\delta$ respectively, not as single curves but, rather, as bands. 
For example, the relative error on $\theta_{13}$ at a given true value of $\theta_{13}$ depends also on the true value of $ \delta$. This weaker dependence 
is shown as an interval corresponding to varying $\delta$ in its full range. The collection of these intervals as a function of $\theta_{13}$ forms what we call a {\it precision band}.
Similarly, the error in $\delta$ is shown at a given true value of $ \delta$ with an interval that represents the variation of this error on the other hidden variable, 
$ \theta_{13}$. 
The range of true values is taken to be the whole physical range for $\delta \in[-\pi, \pi]$, while we choose $\theta_{13}\in [5.7^\circ, 10^\circ]$, 
the lower limit corresponding to the $3\sigma$-range found by Daya Bay, while the upper bound is instead stemming from previous global fits. 

\subsection{Setups}

Regarding the facilities considered, many long-baseline experiments have been proposed to complete the determination of the neutrino mixing parameters, measure the neutrino hierarchy (when sufficiently long baselines are considered) and, hopefully, discover leptonic CP violation. They fall in three main categories: conventional beams and/or super-beams, beta-beams and neutrino factories. 

Super-beams are very intense conventional neutrino beams produced from pion and kaon decays. These beams are mostly composed of muon neutrinos or antineutrinos with an
unavoidable  and non-negligible contamination from other flavours. The appearance of electrons or positrons at a far detector provides a determination of  the oscillation probability $P_{\mu e}$, while the muon disappearance signal gives a precise determination of the atmospheric parameters. Several super-beams have been proposed over the world
in recent years. We will present results for a subset of them:  the LBNE proposal \cite{Akiri:2011dv}; the SPL super-beam from CERN to a water \v Cerenkov detector at Fr\'ejus~\cite{GomezCadenas:2001eua,Campagne:2004wt,Campagne:2006yx,Longhin:2011hn}; a longer baseline option from CERN to a Liquid Argon detector placed at Pyhas\"almi~\cite{Angus:2010sz} (C2P); and  the T2HK proposal \cite{Itow:2001ee,Nakamura:2003hk,Abe:2011ts}. Note that our simulation of the T2HK setup follows the original proposal~\cite{Itow:2001ee,Nakamura:2003hk}. The more recent LOI~\cite{Abe:2011ts} describes a setup with a beam of lower power but a slightly more massive detector and modified fluxes, efficiencies and systematics. 
We find that, despite the modifications in the more recent setup, the performance of the newer version of the facility in Ref.~\cite{Abe:2011ts} is in rough agreement with our simulation of the original proposal. 

Beta-beams are very intense $\nu_e$ or $\bar{\nu}_e$ beams produced from boosted radioactive ion decays\cite{Zucchelli:2002sa}. The beam has 
no other contamination and the flux can be determined with very good accuracy  from $\beta$-decay kinematics, by measuring the parent ion energy. The appearance
of muons in a far detector allows to measure the golden oscillation probability $P_{e\mu}$. Additional information could be obtained from the observation of oscillations in 
the $\nu_e$ disappearance channel (and its CP conjugate). However, this channel is systematics-dominated and turns out to be rather ineffective~\cite{Donini:2004iv}. The absence of $\nu_\mu$ in the 
flux puts the precise measurement of the atmospheric parameters out of reach for beta-beams.  This is a severe limitation of this facility, since we have found that a precise measurement 
of atmospheric parameters is mandatory to achieve a good precision on $\delta$ (see also Ref.~\cite{Donini:2005rn}). For this reason, we will combine the beta-beam simulations 
with information from the disappearance channel at T2K. Such combination is not necessary for the other facilities considered in this paper, since their expected precision in the atmospheric parameters is already expected to be better than that achievable at T2K.

 The spectrum and intensity of a beta-beam flux is fixed by the number of decaying ions, the type of ion 
 ($^6$He, $^{18}$Ne are the preferred choices) and the boost factor $\gamma$. These ions can be boosted up to $\gamma \simeq 150 (250)$ for $^6$He/$^{18}$Ne, 
 respectively, when using the existing facilities at CERN. Replacing the SPS with a new refurbished accelerator would allow to boost the same ions up to 
 $\gamma=350/580$. It has been shown that the beta-beam physics reach improves with $\gamma$, due to their larger neutrino energies and longer baselines~\cite{BurguetCastell:2003vv, BurguetCastell:2005pa}. We have therefore considered both a low-$\gamma$ option~\cite{Mezzetto:2003ub,Donini:2004hu,Mezzetto:2004gs,Donini:2004iv,Donini:2005rn,Huber:2005jk,Campagne:2006yx, FernandezMartinez:2009hb} (produced from the decay of \He/\Ne boosted to $\gamma=100$), and a high-$\gamma$ setup~\cite{BurguetCastell:2003vv, BurguetCastell:2005pa,Agarwalla:2005we,Donini:2006tt,Volpe:2006in,Agarwalla:2006vf,Donini:2007qt,Jansson:2007nm,Agarwalla:2007ai,Coloma:2007nn, Meloni:2008it,Agarwalla:2008gf,Agarwalla:2008ti,Winter:2008cn,Winter:2008dj,Choubey:2009ks,Coloma:2010wa}, produced from the decay of the same ions boosted at $\gamma=350$. 
 These setups will be referred to as BB100 and BB350, respectively. 
 
Neutrino factories are also intense $(\nu_e,\bar{\nu}_\mu)$ or $(\bar{\nu}_e,\nu_{\mu})$ beams resulting from boosted and cooled $\mu^+$ and $\mu^-$ that decay in the straight sections of a storage ring aiming at a far detector.  As in the case of the beta-beam, the neutrino flux is known very accurately, but in contrast with the beta-beam the charge
of the muon in the far detector needs to be determined, because the measurement of the $P_{e\mu}$ comes from the determination of a small
wrong-sign muon component in a large sample of right-sign muons.  It is mandatory, therefore, to have a magnetizable detector for this facility. 
The right-sign muon measurement, on the other hand, gives the muon disappearance probability from which the atmospheric parameters can be precisely determined.  

Until recently, the baseline scenario of recent studies was the IDS-NF \cite{NF:2011aa}. In this scenario, $E_\mu \sim 25$ GeV and two baselines at $4000$ and $7500$~km were considered.  However, the detector placed at 7500 km from the source is mainly needed to solve degeneracies for very small values of $\theta_{13}$ ($\sin^22\theta_{13}\lesssim 10^{-3}$), see Refs.~\cite{BurguetCastell:2001ez,Huber:2003ak}. Therefore, in light of the recent measurements of T2K and Daya Bay, the magic baseline is most probably unnecessary. We will therefore consider a $25$ GeV one-baseline neutrino factory, which we call IDS1b.  On the other hand, a lower-energy neutrino factory~\cite{Geer:2007kn,Bross:2007ts,FernandezMartinez:2010zza}  with $E_\mu=10$ GeV~\cite{Agarwalla:2010hk} has been proposed as optimal if $\theta_{13}$ is large. We will refer to this setup as LENF. 
 
 In App.~\ref{sec:appendix} we provide the details of the beam and detectors for all the setups that have been included in this study. Tab.~\ref{tab:summary} simply summarizes the values of some variables that 
determine to a large extent their physics reach. These are: 
\begin{itemize}
\item the baseline $L$;
\item  the number of signal charged-current events in the assumption of maximal golden channel conversion for both beam polarities, $N_\nu/N_{\bar{\nu}}$, which gives an idea of the  real statistical power of each setup;
\item the number of background events to the golden signal, $B_{\nu}/B_{\bar{\nu}}$;
\item the average of the neutrino or antineutrino energy, $\langle E_\nu\rangle/ \langle E_{\bar{\nu}}\rangle$, of the fully converted events\footnote{Note that, for some facilities, this average value is sometimes higher than the one required to be at the first oscillation peak, which translates in a poorer performance. 
This is, for instance, the case for the SPL setup, with a very high number of events at a mean energy of $0.58$~GeV, 
far from the oscillation peak at the 130~km baseline at $\sim 0.26$~GeV.};
\item the dispersion of the  neutrino/antineutrino energy, $\delta E_\nu/ \delta E_{\bar{\nu}}$, which gives an idea of the wideness of the beam;
\item the average strength of matter effects, defined by
\bea
\hat{A} \equiv {2  \sqrt{2} \langle E_\nu \rangle G_F n_e  \over  |\Delta m^2_{13}| }.
\eea
\end{itemize}
 
As it is clear from the table all beams are rather wide, the narrowest being the off-axis flux of T2HK with a spread of $\sim$22$\%$, while the widest
is LBNE with $\sim$39$\%$.  Statistics is more significant in the neutrino factory setups, followed by the short-baseline super-beams (T2HK, SPL). 
The more statistically limited setups are the long-baseline super-beams and the beta-beams. Backgrounds are more significant
in T2HK, while they are almost negligible for the neutrino factories. Finally, matter effects are largest for the neutrino factory setups, followed by the long-baseline super-beams.
 
\begin{table}[htb]
\begin{center}{
\renewcommand{\arraystretch}{1.6}
 \begin{tabular}{c|cccccc}
   & $L$ & $N_\nu/N_{\bar{\nu}}$  & $B_\nu/B_{\bar{\nu}}$  & $\langle E_\nu \rangle/\langle E_{\bar{\nu}} \rangle$ & $\delta E_\nu/\delta E_{\bar{\nu}}$  & $\hat{A}$  \\ \hline \hline
T2K & 295 & 2.6/0 $\times 10^3$ & 46/0 & 0.72/-- & 0.27/-- & 0.02 \\ \hline
NO$\nu$A & 810 & 1.1/0.7 $\times 10^3$ & 10/11 & 2.02/2.04 & 0.43/0.42 & 0.14 \\ \hline \hline
T2HK & 295 & 4.3/1.3 $\times10^5$ & 4.3/1.5 $\times 10^3$ & 0.79/0.80 & 0.18/0.18 & 0.022\\ \hline 
LBNE & 1290 & 2.3/0.9 $\times 10^4$ & 302/201 & 3.55/3.50 & 1.38/1.33 & 0.30\\ \hline
SPL & 130 & 2.5/1.6 $\times 10^5$ & 1.1/1.2 $\times 10^3$ & 0.59/0.57 & 0.20/0.21 & 0.017\\ \hline 
C2P & 2300 & 2.4/1.1 $\times 10^4$  & 210/129 & 5.04/5.15 & 1.65/1.59 &  0.48\\ \hline\hline
BB100 & 130 & 2.9/4.4 $\times 10^4$ & 0.6/1.2 $\times 10^3$ & 0.47/0.45 & 0.18/0.18 & 0.013\\ \hline
BB350 & 650 & 5.0/9.2 $\times 10^4$ & 372/432 & 1.53/1.61 & 0.45/0.45 & 0.11\\ \hline\hline
LENF & 2000 & 8.1/5.3 $\times 10^5$ & 48/81 & 6.75/6.78 & 1.81/1.79 & 0.63\\ \hline
IDS1b &4000 & 1.9/1.2 $\times 10^6$ & 154/196 & 16.85/16.86 & 4.57/4.55 & 1.65\\ \hline\hline
 \end{tabular}}
\caption{Summary of the main details of the setups considered. From left to right the columns present: the experiments baseline (in km); the total number of signal neutrinos and antineutrinos including detector efficiencies and assuming a full flavour conversion of all events; the total number of background events for the neutrino and antineutrino channels; 
the mean true energy of the total events (in GeV); the energy dispersion of the total events (in GeV); and, the size of the matter effects parametrized as $\hat{A}$.  
\label{tab:summary} }
\end{center}
\end{table}

\section{Precision on $\theta_{13}$ and $\delta$}
\label{sec:precision}

In this section, we derive simple analytical arguments that allow to understand the basic features of the results of Sec.~\ref{sec:results}. 
In particular, we are interested in understanding the dependence of precision  on the true values of $\theta_{13}$ and $\delta$. 
We first consider the approximate\footnote{In the case of large $\theta_{13}$, a more accurate expansion of the probability can be found in Ref.~\cite{Asano:2011nj}. We consider here the approximate probability expanded in the assumption of small $\theta_{13}$ only for illustration purposes, while for the numerical results presented throughout this paper the exact probabilities are used instead.} golden channel probability~\cite{Cervera:2000kp,Freund:2001ui,Akhmedov:2004ny}:
\begin{eqnarray}
P^\pm_{e\mu}(\theta_{13},\delta)  &=&   \theta_{13}^2 ~s_{23}^2 {\sin^2 [ (1\mp \hat{A}) \Delta ]  \over (1\mp \hat{A})^2} + c_{23}^2  \sin^2 2 \theta_{12} {\Delta^2_{12} } \left[{\sin(\hat{A} \Delta) \over \hat{A}\Delta}\right]^2 \nonumber\\
&+&  \theta_{13} ~Ê 2 \sin 2 \theta_{12} \sin 2\theta_{23}  {\Delta_{12} \over \Delta} \cos (\Delta \mp \delta) {\sin(\hat{A} \Delta) \over \hat{A}} {\sin[(1\mp \hat{A}) \Delta] \over 1\mp \hat{A}}, \nonumber\\
&=& P^\pm_{\mu e}(\theta_{13},-\delta) \, , 
\end{eqnarray}
where 
\bea
\Delta \equiv {\Delta m^2_{13} L\over 4 E},~Ê~Ê \Delta_{12} \equiv {\Delta m^2_{12} L\over 4 E} ,~Ê \hat{A} \equiv { \sqrt{2} G_F n_e L \over 2 \Delta},  
\eea
and the $\pm$ corresponds to neutrino or antineutrinos. $L$ and $E$ are the baseline and neutrino/antineutrino energy, respectively. 
Although the number of events $N$ in a given channel corresponds to the convolution of the probability with neutrino fluxes, 
$\nu N$ CC cross-sections and detector efficiencies, we will see that the dependence of $\Delta \theta_{13}$ and $\Delta \delta$ on the true values of 
$\theta_{13}$ and $\delta$ can be understood assuming that $N \propto P$. 

\subsection{Precision on $\theta_{13}$}

We assume that all the considered facilities will measure two CP conjugated channels and that the considered detectors will provide several energy bins. 
We also assume that these measurements allow the determination of $\theta_{13}$ and $\delta$ with negligible correlations (this is a reasonable assumption as long as 
 the intrinsic degeneracy \cite{BurguetCastell:2001ez} is solved, as it is the case for most of the facilities under study in this paper). 
 
 Using standard error propagation
\be
\Delta N_\pm =  \left|{\partial N_\pm \over \partial {\theta_{13}}} \right |_{(\theta_{13},\delta)} (\Delta\theta_{13})_\pm \propto \theta_{13}  {\sin^2 [ (1\mp \hat{A}) \Delta ]  \over (1\mp \hat{A})^2}  (\Delta\theta_{13})_\pm + \dots,
\label{eq:deltat13}
\ee
where we have neglected subleading terms in $P^\pm$.  The error on the weighted average of neutrino and antineutrino data is  
\bea
\Delta \theta_{13} \simeq \left(\sqrt{{1\over (\Delta\theta_{13})^2_+} + {1\over (\Delta\theta_{13})^2_-} }\right) ^{-1}.
\label{eq:t13comb}
\eea
If the error on the number of oscillated events is dominated by the (gaussian) statistical error, then $\Delta N_\pm \simeq \sqrt{N_\pm}$, 
and it follows that $\Delta \theta_{13}$ is approximately independent of $\theta_{13}$ and $\delta$. In this case, 
the relative error decreases linearly with ${\theta}_{13}$: 
\bea
\Delta_r \theta_{13}  \propto {1 \over \theta_{13}} . 
\eea
On the other hand, if the error is dominated by the systematic error on the signal, $\Delta N_\pm \propto N_\pm$, we get $\Delta \theta_{13} \propto \theta_{13}$ and
the relative error on $\theta_{13}$ is independent of $\theta_{13}$.
Finally, if the error is dominated by the error on the background (assumed independent on $\theta_{13}$) $\Delta \theta_{13} \propto 1/\theta_{13}$ and
$\Delta_r \theta_{13} \propto 1/ \theta^2_{13}$. The dependence on $\delta$ in any case is expected to be small.

It is also interesting to understand what is the impact of matter effects on the precision in $\theta_{13}$. From Eq.~(\ref{eq:deltat13}) is is clear that  the minimal error is obtained in the energy bin that maximizes the oscillation term. In the presence of matter, this occurs at different bins for neutrinos and antineutrinos: $\Delta = {\pi \over2} (1\mp \hat{A})^{-1}$. Combining the corresponding optimal errors as in Eq.~(\ref{eq:t13comb}), we find that the error decreases with increasing $\hat{A}$. Essentially, one of the errors for neutrinos or antineutrinos (depending on the sign
of $\hat{A}$) improves and the other worsens, but  the combination always improves. All other conditions being the same, larger matter effects improve the precision in $\theta_{13}$. 

\subsection{Precision on $\delta$ }

Let us now consider the error on $\delta$. Under the same assumptions as before we have in this case
 \bea
\Delta N_\pm \simeq \sqrt{N_\pm} =   \left|{\partial N_\pm \over \partial{ \delta}}\right| (\Delta\delta)_\pm, \label{eq:DeltaN}
\eea
and 
\bea
\Delta \delta \simeq \left(\sqrt{{1\over (\Delta\delta)^2_+} + {1\over (\Delta\delta)^2_-} }\right) ^{-1}. \label{eq:quadrature}
\eea
Now, the dependence on $\delta$ is much less trivial. We find
 \bea
(\Delta \delta)_\pm  \propto \left|~ {\hat{A} \Delta \over \sin \hat{A} \Delta}~Ê {1\over \sin \Delta \mp { \delta}} \right|,
\label{eq:deltadeltapm}
\eea
where we just show the dependence on the parameters of interest. Note in particular that the relative weight of the neutrino and antineutrino error can be different due to the different fluxes and cross-sections. We have assumed that the flux $\times$ cross-section goes as $E^2$. Deviations from this behaviour will be different for the different setups under consideration. We will therefore ignore these effects for the time being and indicate where they could make a difference. 

As it can be seen from Eq.~(\ref{eq:deltadeltapm}), the error now depends non trivially also on $L/E$. 
Let us now consider various situations.

\subparagraph{Vacuum}

In this case the oscillation probability  is maximal for neutrinos and antineutrinos at the same $L/E$, corresponding to the condition $\Delta = (2 n+1) \pi/2$, with $n$ integer. Let us suppose that we have a narrow beam at the $L/E$  corresponding to the first oscillation maximum. We have then 
\bea
(\Delta \delta)_\pm \propto {1\over \sin\left( {\pi\over 2} \mp \delta\right)}.
\eea
The combination of neutrinos and antineutrinos gives
\bea
\Delta \delta \propto \sqrt{{1\over 1+\cos 2 \delta}}. \label{eq:deltavac}
\eea
Thus, the error has a very strong dependence on $\delta$ which actually diverges if $\delta \rightarrow {\pi\over 2}, {3 \pi \over 2}$. 

\begin{figure}[htb!]
\begin{center}
\includegraphics[width=7cm]{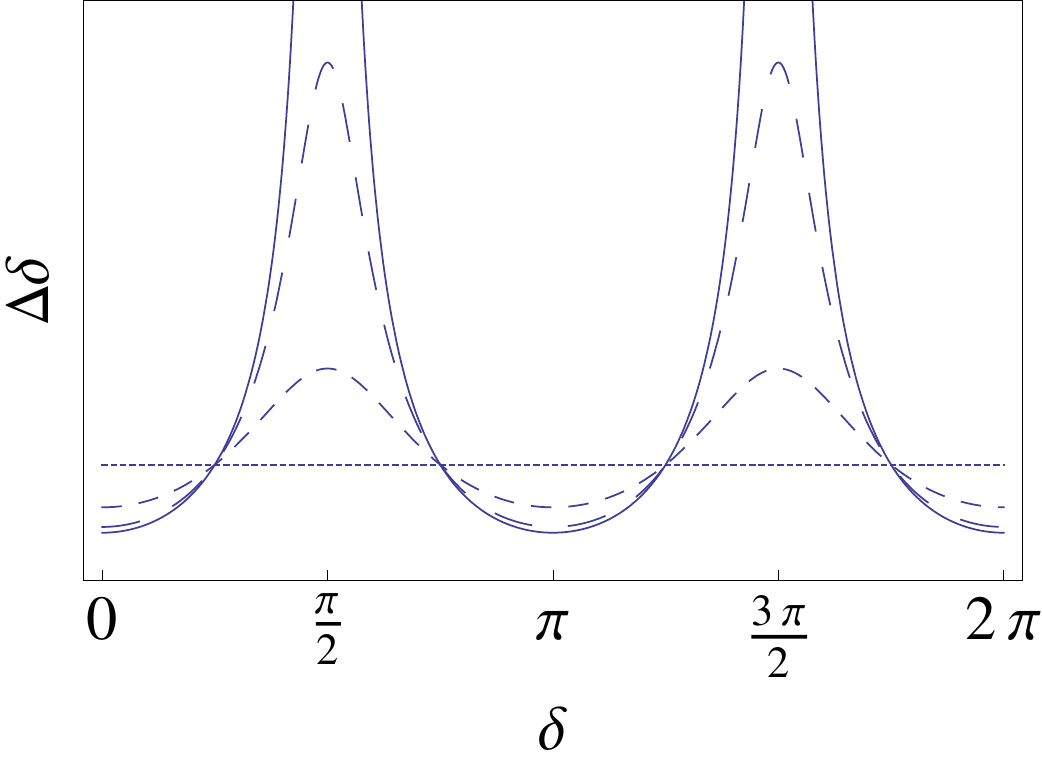} \includegraphics[width=7cm]{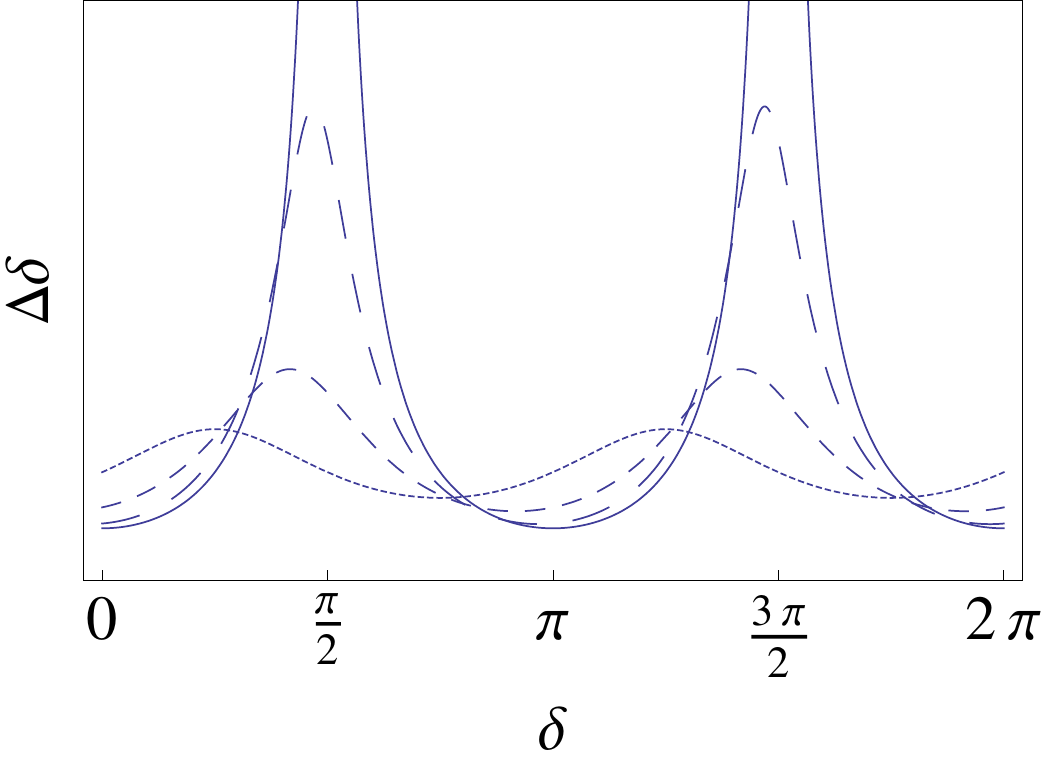}
\caption{ $\Delta\delta$ as a function of $\delta$ for $\Delta={1\over 2}, {2\over 3}, {5\over 6},1 \times {\pi\over 2}$ (dotted to solid)  and with negligible matter effects, assuming the same weight for neutrinos and antineutrinos (left) or 50$\%$ less antineutrinos (right). The error plotted here corresponds to the approximate formula in Eq.~(\ref{eq:deltavac}).}
\label{fig:delta_vac}
\end{center}
\end{figure}

If we move away from the oscillation maximum both to higher or lower values of $L/E$ the dependence on $\delta$ smooths out. On the left plot of  Fig.~\ref{fig:delta_vac} we show the result for $\Delta = ({1\over 2}, {2\over 3}, {5\over 6}, 1) \times {\pi\over 2}$, assuming the same weight for neutrinos and antineutrinos. We see that the error is constant only for some values 
of  $\Delta =\pi/4, 3 \pi/4,..$, while it is maximal at $\delta= \pi/2, 3 \pi/2$ and minimal at $\delta = 0, \pi$. 
The best error bar is smaller when the experiment is close to oscillation maxima, but the worst error is also largest at the same point. 
This indicates that if we just look at the sensitivity to CP violation we would rather be at $\Delta=\pi/2$, but if we instead look at the average precision on $\delta$ for any $\delta$ that is not the preferred situation. 

Clearly this also shows that those neutrino beams that in practice provide sufficient information outside the peak, {\em i.e.} sufficient 
energy dependence, can help to reduce the variation of $\Delta\delta$ with $\delta$.

When the weight for neutrinos and antineutrinos is significantly different and we consider bins outside the peak, the maxima of $\Delta\delta$ shift to the left (right) in $\delta$, if the fraction of antineutrinos is less (more) than that of neutrinos. This is shown in the right panel in Fig.~\ref{fig:delta_vac}. Also the error is no longer constant for the special values
$\Delta=\pi/4, 3 \pi/4$. 

\subparagraph{Matter}
 
In matter, the maxima of the oscillation probability for neutrinos and antineutrinos do not coincide. It is
sensible to assume  that most of the information in the neutrino channel comes  from the bin where the  neutrino probability maximizes, 
 {\em i.e.} $(1-\hat{A}) \Delta = \pi/2$,  while in the antineutrino channel  it comes from the bin where the antineutrino probability  maximizes,  {\em i.e.} $( 1+ \hat{A}) \Delta = \pi/2$. The contribution to the error of both such bins is
\bea
(\Delta\delta)_\pm =  {{\pi \over 2}~{\hat{A}\over (1\mp \hat{A})} \over \sin \left[{\pi\over 2} {\hat{A}\over (1\mp\hat{A})}\right]} {1 \over \sin\left({\pi \over 2}~{1 \over (1\mp \hat{A})} \mp \delta\right)}, \label{eq:deltamatter}
\eea
while for the T-conjugated channel $\nu_\mu \rightarrow \nu_e$ we must substitute $\delta\rightarrow -\delta$.

\begin{figure}[htb!]
\begin{center}
\includegraphics[width=10cm]{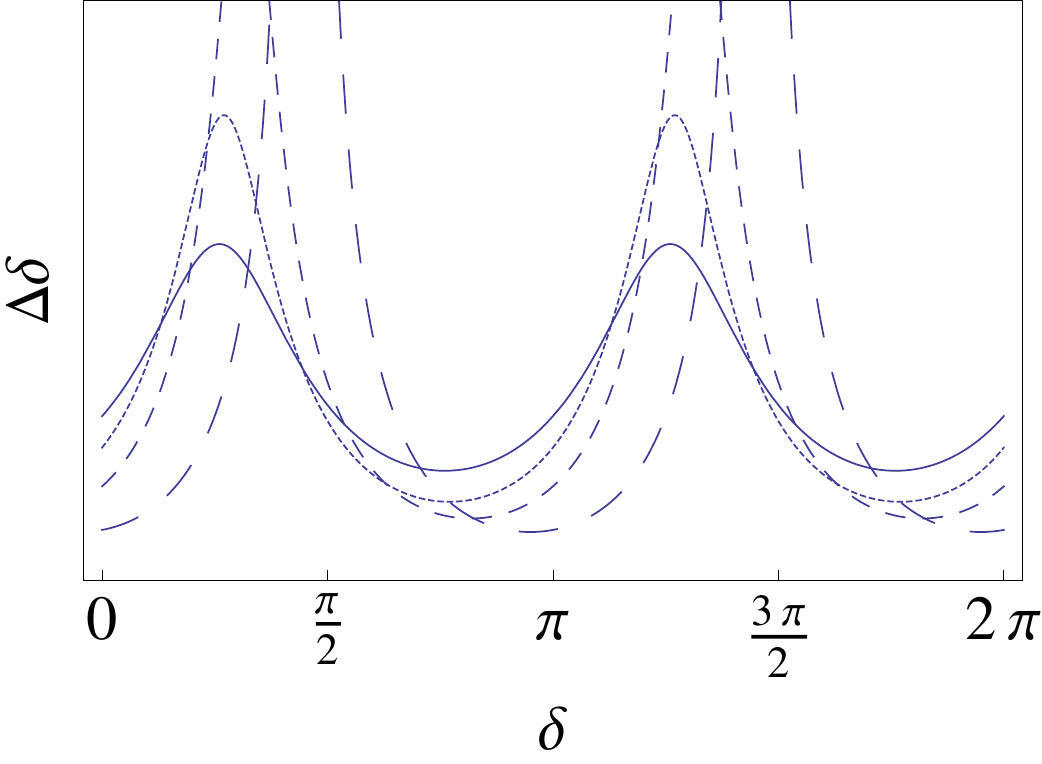}
\caption{ $\Delta\delta$ as a function of $\delta$ for the $\mu \rightarrow e$ channel for $\hat{A}=1/10,1/3,5/12,1/2$ (dashed to solid), assuming the same weight for neutrinos and antineutrinos. The error plotted here corresponds to the result of substituting the approximate formula~(\ref{eq:deltamatter}) in Eq.~(\ref{eq:quadrature}).}
\label{fig:delta_mat}
\end{center}
\end{figure}

Fig.~\ref{fig:delta_mat} shows the dependence of $\Delta\delta$ on $\delta$ for several values of $\hat{A}$, that is, of the strength
of matter effects. 
In this plot we observe the two main implications of matter effects. First the peaks in $\Delta\delta$ move to the left (right) in the $\mu \rightarrow e$ $(e\rightarrow \mu)$ channels. Second the dependence on $\delta$ is smoothed out, but the best achievable precision 
gets worse, as expected since matter effects tend to hide genuine CP violation. 

In this case, if we move away from the peak, that is, $(1\mp\hat{A}) \Delta = \epsilon {\pi\over 2 }$ with $\epsilon \leq 1$, there is no improvement  in $\Delta\delta$. Therefore we expect that energy dependence in the scenarios with large matter
effects will not be so important as in vacuum. 

Under the previous assumptions we do not expect a significant dependence of $\Delta\delta$ on 
$\theta_{13}$ . On the other hand, if the error on $N_\pm$ is not dominated by the statistical error but by the systematics of the signal or the  background, a similar analysis shows that we should expect some dependence on $\theta_{13}$. In particular, when the systematics on the background dominates, we expect that the error is inversely proportional to $\theta_{13}$. 

\begin{figure}[htb!]
\begin{center}
\includegraphics[width=15cm]{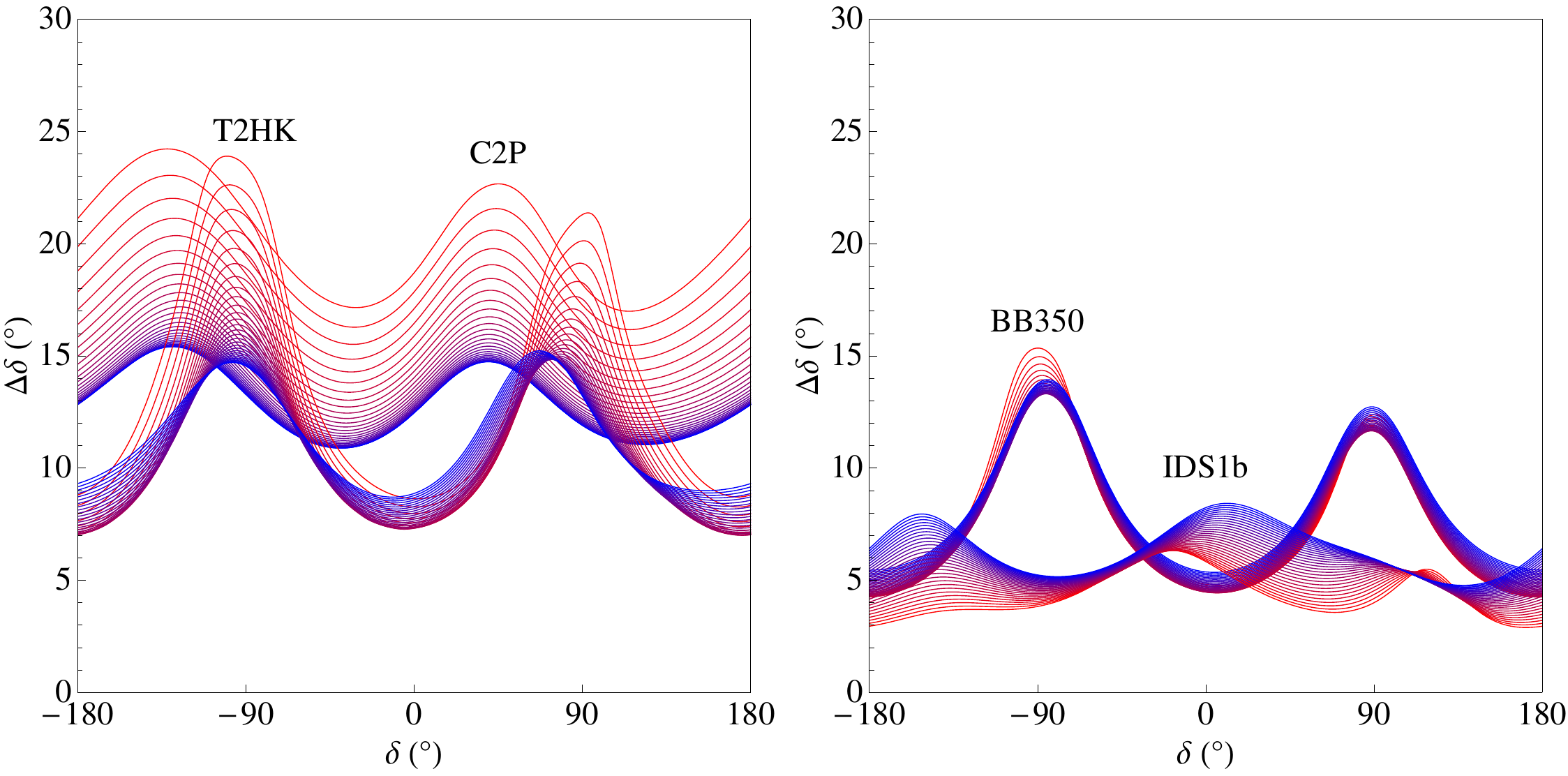}
\caption{ $\Delta\delta$ as a function of $\delta$ for the $\mu \rightarrow e$ channel (left) in the C2P and T2HK setups, 
and for the $e\rightarrow \mu$ channel (right) in the BB350 and IDS1b setups. The different curves from red to blue correspond to
different values of $\theta_{13}=3-10^\circ$, in steps of $0.25$ degrees.  }
\label{fig:rojoazul_vac}
\end{center}
\end{figure}

This naive analysis seems to explain rather well the qualitative features of  the precision on $\delta$ found in all scenarios considered in this paper. $\Delta \delta$ as a function of $\delta$  is shown in Fig.~\ref{fig:rojoazul_vac} for four facilities: T2HK and the BB350, which both involve small matter effects, falling therefore in the vacuum category; and the C2P and the IDS1b, that  involve a significantly longer baseline and very significant matter effects (see Tab.~\ref{tab:summary}). The different curves correspond to different values of $\theta_{13}=3-10^\circ$ (red to blue).  As expected the maxima move to the left for the long-baseline super-beam with respect to the short-baseline one, while they move to the right for IDS1b with respect to the BB350. 

The degradation of the error is very significant for the super-beams for $\theta_{13}$ below $6^\circ$. As explained above, the naive expectation is that, if the error is statistically dominated, it should not  depend on $\theta_{13}$. A dependence is expected however when the error becomes systematics dominated. If it is due to the background systematics we expect a degradation with decreasing $\theta_{13}$, while if it is due to signal systematics we expect a degradation with increasing $\theta_{13}$.  Indeed the degradation in the super-beams agrees with the expectation of being background-systematics dominated.  
The situation for the BB350 and IDS1b setups is different. These are much purer beams so background systematics start to be significant only for smaller values of $\theta_{13}$. The error for the BB350 indeed shows a degradation at significantly smaller values of $\theta_{13}$ compared to the super-beams.  The IDS1b the error appears consistent with being dominated by the systematics on the signal, therefore improving with decreasing $\theta_{13}$.


\section{Results}
\label{sec:results}

We proceed now to compare the different setups\footnote{ 
Note that the characteristics of each of the considered setups (baseline, energy, flux, efficiencies, backgrounds and systematic errors) are fixed and 
that we have not studied how changing some or all of them affects the precision observables. Our results would change if, for instance, the systematic errors 
or the neutrino fluxes of a given experiment are varied. } 
on the basis of the precision observables defined in Sec.~\ref{sec:setups}.

Solar and atmospheric input parameters have been fixed to their present best fit results from the global fit analysis in
Ref.~\cite{Schwetz:2011qt}: $\theta_{12}=34.2^\circ$, $\Delta m^2_{12}=7.64\times 10^{-5}$ eV$^2$,  $\theta_{23}=45^\circ$, $\Delta m^2_{31}=2.45\times 10^{-3}$ eV$^2$.
A normal hierarchy has been assumed in all cases. 
The confidence regions have been obtained after marginalization over solar and atmospheric parameters, 
assuming the following $1 \sigma$ gaussian priors: $3 \%$ for $\theta_{12}$, $2.5 \%$ for $\Delta m^2_{12}$, $8 \%$ for $\theta_{23}$ and $4 \%$ for $\Delta m^2_{31}$. Finally, a $2\%$ uncertainty over the PREM density profile~\cite{prem} has been also considered. Note that both super-beams and neutrino factories are sensitive to the disappearance channel. Therefore, all of them are going to improve over the priors listed above. This is not the case for the beta-beams, though, for which the data from T2K would provide the effective priors for the atmospheric parameters instead.

As already stressed, no sign degeneracies have been considered for the results presented here. For the region of the $\theta_{13}$ parameter space allowed by the Daya Bay data \cite{An:2012eh},  the neutrino factories and most of 
the beta-beam and super-beam setups considered are able to measure the hierarchy. This is very unlikely, however, for T2HK, the SPL and the $\gamma=100$ He/Ne beta-beam. There is some possibility that these experiments could measure the hierarchy through atmospheric neutrino data, though (see, for instance, Refs.~\cite{Huber:2005ep,Campagne:2006yx}), or from their combination with INO and/or NO$\nu$A data (see, for instance, Refs.~\cite{TabarellideFatis:2002ni,Bernabeu:2003yp,PalomaresRuiz:2004tk,Indumathi:2004kd,Petcov:2005rv,Arumugam:2005nt,Samanta:2006sj,Gandhi:2007td,Mena:2008rh,Gandhi:2008zs,Samanta:2009qw,Blennow:2012gj}). For the results presented here a normal hierarchy is always assumed. We have checked that the results for an inverted hierarchy are very similar, although slightly deteriorated for the facilities with lower antineutrino (vs neutrino) rates at the detector.

\subsection{Precision on $\theta_{13}$ }
\label{sec:th13prec}

We first study the precision on $\theta_{13}$ that can be attained at the considered setups. It is interesting to see how the precision of future facilities on $\theta_{13}$ would compare to the achievable precision at reactor experiments and, in particular, to the precision achievable at Daya Bay. The present error at the $1\sigma$ CL on $\theta_{13}$ is $ 9.3\%$, after only 55 days of data taking. We show this result as an empty triangle in the figures. In view of this, it seems reasonable to assume that this error will eventually be improved down to the systematic level. Assuming that the best fit does not change in the future, this would correspond to a relative precision of $2.8\%$, which is indicated through the black stars in the figures. We have also included a vertical line at $ \theta_{13} = 5.8^\circ$, which corresponds to the 3$\sigma$ lower bound on $\theta_{13}$ for the Daya Bay result. 

\begin{figure}[htb!]
\begin{center}
	\includegraphics[width=15cm]{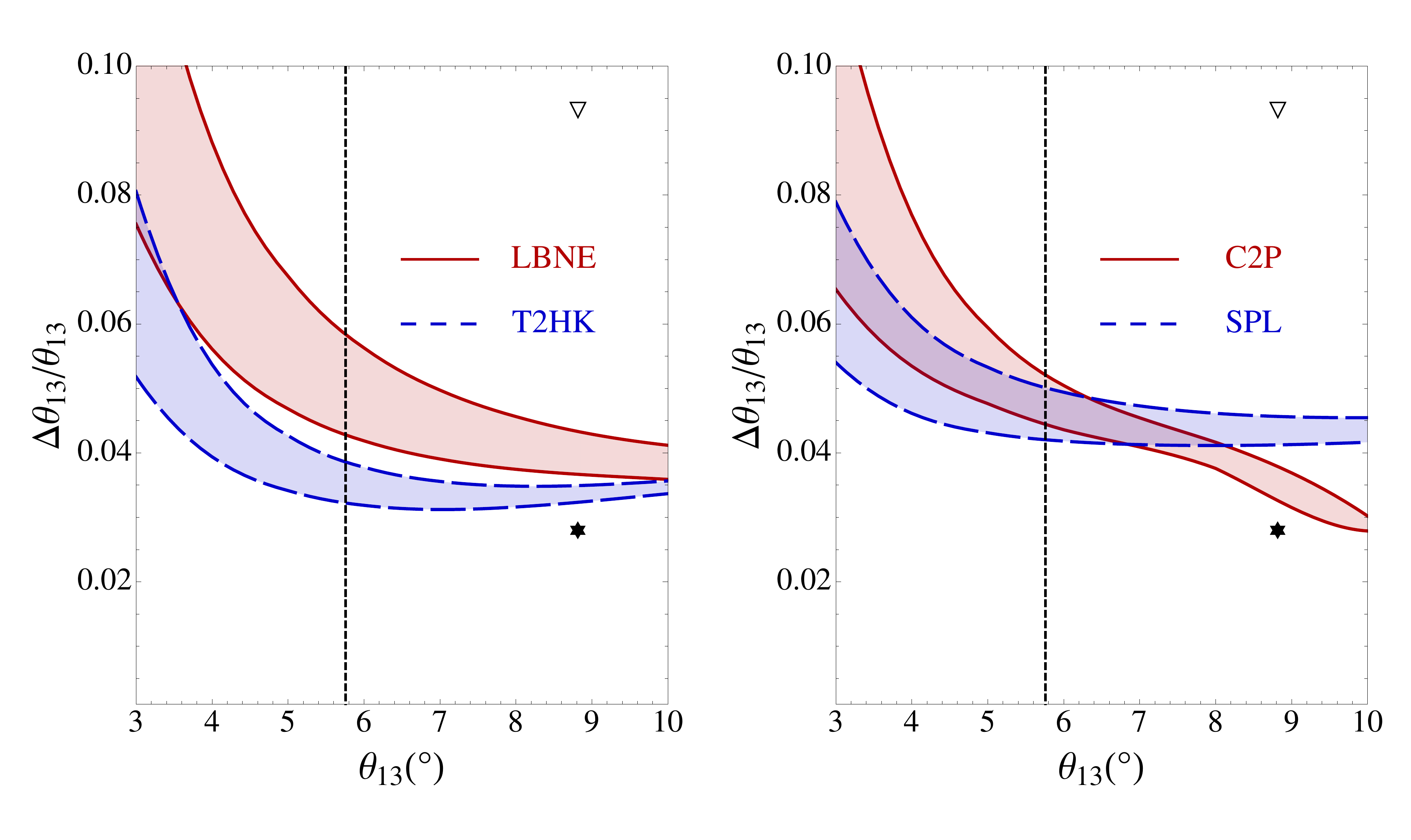}

\caption{Relative error on $\theta_{13}$ as a function of $ \theta_{13}$ at $1 \sigma$ (1 d.o.f.) for the considered super-beam setups. 
Left panel: results for T2HK (blue, dashed lines) and LBNE (red, solid lines). 
Right panel: results for SPL (blue, dashed lines) and C2P (red, solid lines). The width of the bands shows the dependence with the value of $\delta$. The empty triangle shows the present precision at $1\sigma$ for Daya Bay, while the star represents the ultimate attainable precision, corresponding only to the quoted systematic error.  Both points are shown for the present best fit. The vertical line corresponds to the present Daya Bay 3$\sigma$ lower bound. 
A true normal hierarchy has been assumed and no sign degeneracies have been taken into account. 
 }
\label{fig:compTh13RelSB}
\end{center}
\end{figure}

In Fig.~\ref{fig:compTh13RelSB} we present a comparison in terms of precision for the super-beam setups defined in Sec.~\ref{sec:setups}. The comparison of both panels indicate that, within the Daya bay 3$\sigma$ region, all facilities have a comparable performance reaching a precision below $\sim 5\%$. T2HK performs slightly better, with a precision below 4$\%$ in the whole region. It is remarkable, however, that none of the considered super-beams can improve over the expected ultimate precision of Daya Bay. 

Within the Daya bay 3$\sigma$ region, we can see that the scaling with $\theta_{13}$ of $\Delta_r \theta_{13}$ 
of ``short" (T2HK and the SPL) and ``long" (LBNE and C2P) baseline super-beams is different: for short baseline super-beams,  the relative precision on $\theta_{13}$
is roughly independent of ${\theta}_{13}$, indicating that precision in these facilities is limited by the systematics of the signal in this regime; 
for long baseline super-beams the precision improves with ${\theta}_{13}$, instead, as expected when the error is statistics-dominated. Below the Daya Bay 3$\sigma$ bound, 
on the other hand, all super-beams show a significant degradation of $\Delta_r \theta_{13}$. This is due to the fact that, for such small values of $\theta_{13}$, the signal is considerably reduced and the systematics on the background start to dominate the error instead.
The bands are in all cases relatively narrow, which means that the precision on $\theta_{13}$ does not depend significantly on ${\delta}$. 

\begin{figure}[htb!]
\begin{center}
	\includegraphics[width=15cm]{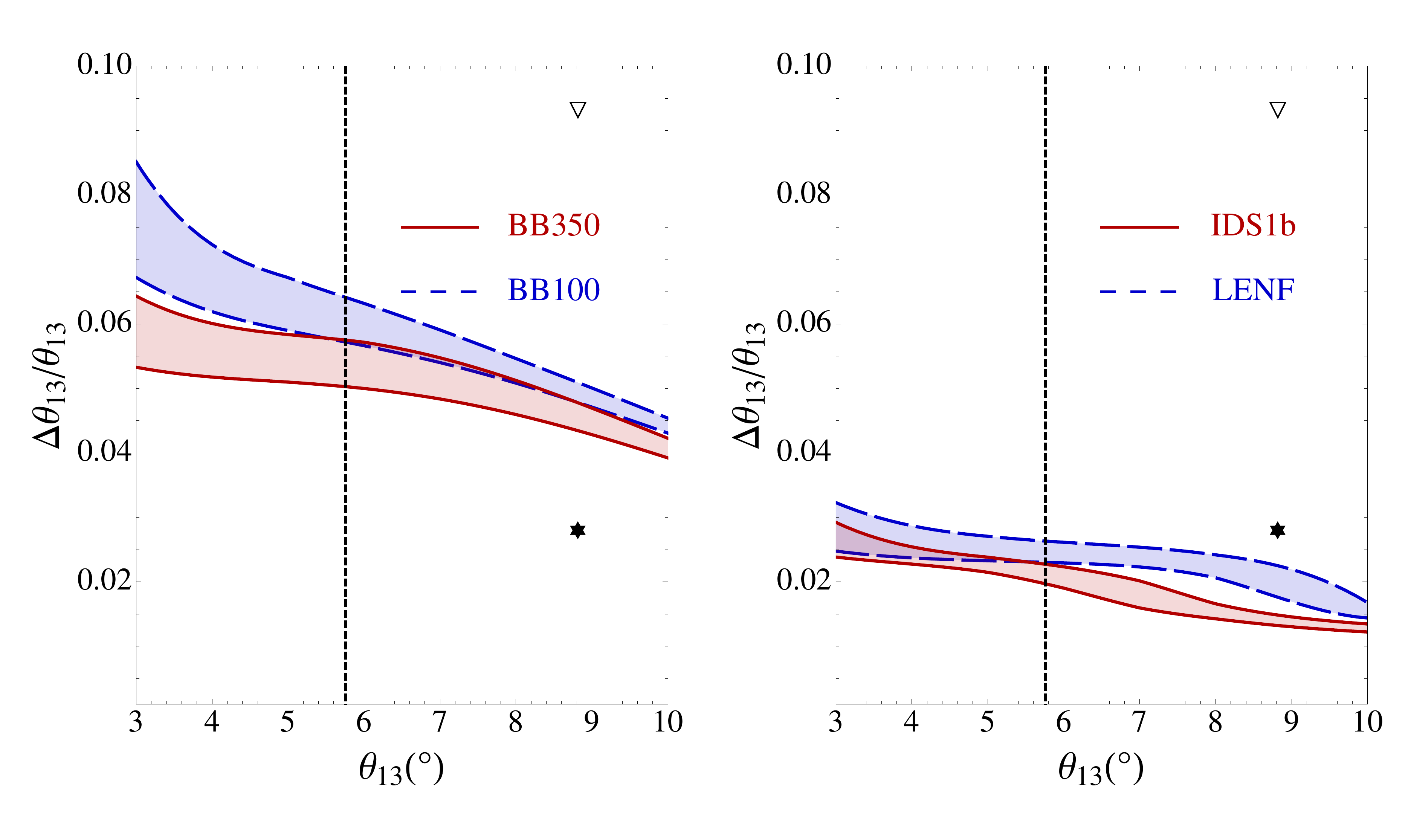}

\caption{Relative error on $\theta_{13}$ as a function of $ \theta_{13}$ at $1 \sigma$ (1 d.o.f.) at the considered beta-beam (left) and neutrino factory (right) setups. 
Left panel: results for BB100 (blue, dashed lines) and BB350 (red, solid lines). 
Right panel: results for LENF (blue, dashed lines) and IDS1b (red, solid lines). 
The width of the bands shows the dependence with the value of $\delta$. The empty triangle shows the present precision at $1\sigma$ for Daya Bay, while the star represents the ultimate attainable precision, corresponding only to the quoted systematic error.  Both points are shown for the present best fit.  The vertical line corresponds to the present Daya Bay 3$\sigma$ lower bound. 
A true normal hierarchy has been assumed and no sign degeneracies have been taken into account.  
 }
\label{fig:compTh13RelBBNF}
\end{center}
\end{figure}

In Fig.~\ref{fig:compTh13RelBBNF} we compare the precision on $\theta_{13}$ attainable in the beta-beam and neutrino factory setups. For all of these setups we can see that 
the precision improves linearly with ${\theta}_{13}$, indicating that $\Delta \theta_{13}$ is statistically dominated. This is not surprising, 
since backgrounds and systematic errors are typically under better control at beta-beam and neutrino factory facilities with respect to super-beams. 

The attainable precision on $\theta_{13}$ at both beta-beam setups ranges from $\sim 6$\% to $\sim 4$\% within the Daya Bay 3$\sigma$ allowed region. 
This is significantly worse than the performance of the considered super-beams, even though the expected number of unoscillated events at the detector is larger 
than for the LBNE and C2P proposals (see Tab.~\ref{tab:summary}). This is because matter effects are small for the beta-beam, while they are very significant in the LBNE and C2P setups. As shown in Sec.~\ref{sec:precision}, matter effects are helpful to reduce the error on $\theta_{13}$.  In the case of the SPL and T2HK setups, even though matter effects are also small, the larger statistics compensates resulting in a similarly good measurement of $\theta_{13}$. The relative performance of super-beam and beta-beam setups is, however, of little relevance, considering that none of these setups can improve over a systematics-dominated measurement by Daya Bay. 

Only the neutrino factories could reduce the $1 \sigma$ range below $\sim 2\%$. We can indeed see that,  within the 3$\sigma$ Daya Bay region, 
the IDS1b (LENF) setup reaches a relative precision on $\theta_{13}$ that ranges from 2.5\% (2.7\%) to 1.4\% (1.5\%). Both facilities
outperform significantly the considered super-beams and beta-beams. 

\subsection{Precision on $\delta$ }
\label{sec:deltaprec}

We now consider  the precision on $\delta$ that can be attained at the considered setups. 

\begin{figure}[htb!]
\begin{center}
	\includegraphics[width=7cm]{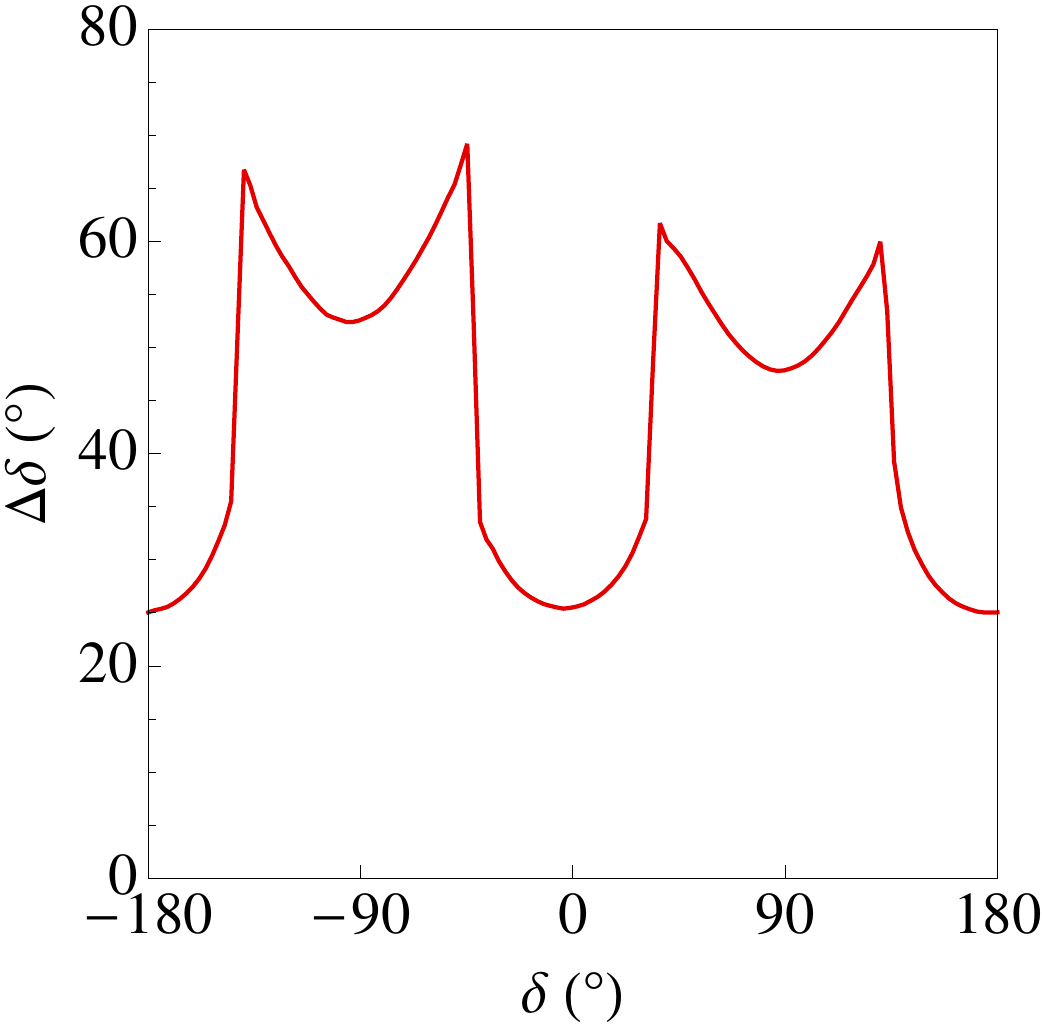}
\caption{Error on $\delta$ as a function of $ \delta$ at $1 \sigma$ (1 d.o.f.) for the combination of T2K and NO$\nu$A, for $\theta_{13} = 8.8^\circ$. A true normal hierarchy has been assumed, and no sign degeneracies have been taken into account. 
 }
\label{fig:deltat2knova}
\end{center}
\end{figure}

First of all, we show in Fig.~\ref{fig:deltat2knova} the error on $\delta$ as a function of the true value of $\delta$ for the combination of T2K and NO$\nu$A, for $\theta_{13} = 8.8^\circ$.
Note that the error on $\delta$ is larger than $25^\circ$ for any value of $\delta$, {\em i.e.} the precision on $\delta$ of these facilities is rather poor.
As can be seen from the figure, two large peaks appear for $\delta \sim \pm \pi/2$, as expected from the analytical results in Sec.~\ref{sec:precision}. The fine structure of the 
peaks is due to the intrinsic degeneracies. The intrinsic degeneracy location strongly depends on the energy and on the beam characteristics \cite{Donini:2003vz}. In vacuum, it appears roughly at the same value of $\theta_{13}$ of the true solution but with $\delta$ shifted to $\pi - \delta$~\cite{BurguetCastell:2001ez}. The true solution and the intrinsic degeneracy become very close around $\delta = \pm \pi/2$ and eventually fuse into a single region, which can be hard to resolve with insufficient energy resolution. The double peak structure seen in Fig.~\ref{fig:deltat2knova} around $\delta = \pm \pi/2$ corresponds to the point in which the intrinsic degeneracy appears and joins with the true solution, dramatically increasing the error on $\delta$. When $\delta$ is exactly $\pm \pi/2$, the true solution and its intrinsic degeneracy overlap and a local minimum appears with better precision.

In Fig.~\ref{fig:compDeltaRelSB} we present a comparison of  the super-beam setups. The qualitative behaviour described in  Sec.~\ref{sec:precision} is clearly observed in the numerical results.  The short-baseline super-beams, T2HK and SPL, fall into the {\it vacuum } category, and attain the worst (best) precision at ${\delta}= \pm \pi/2$ ($\delta = 0,\pi$). The longer baseline super-beams, LBNE and C2P, for which matter effects are very significant (seeTab.~\ref{tab:summary}) fall into the {{\it matter} category 
and achieve the worst precision at smaller values of ${\delta}$. The shift of the positions of maxima and minima is more significant in the case of the longer baseline, C2P. 
Both facilities in the {\it vacuum} regime have similar results, with $\Delta \delta$ ranging from $7^\circ$ ($8^\circ$) to $16^\circ$ ($15^\circ$) at T2HK (the SPL). 
On the other hand, C2P presents the best performance in the {\it matter} regime, with $\Delta \delta$ ranging from $11^\circ$ to $17^\circ$. 
The dependence on ${\theta}_{13}$, {\em i.e.} the width of the bands, is rather small and it will be further reduced as the error on $\theta_{13}$ will improve. 

\begin{figure}[htb!]
\begin{center}
	\includegraphics[width=15cm]{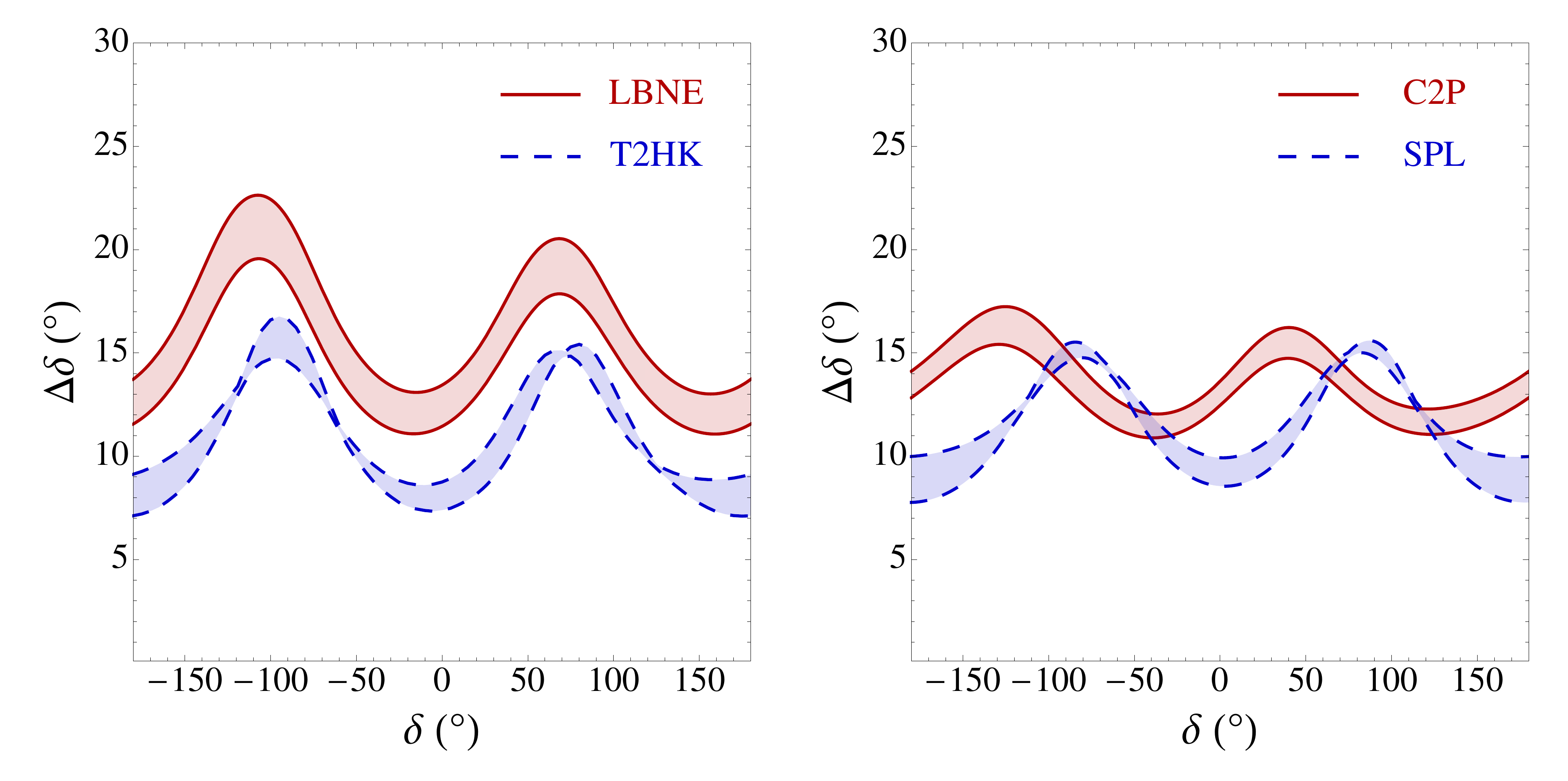}
\caption{Error on $\delta$ as a function of $ \delta$ at $1 \sigma$ (1 d.o.f.) at the considered super-beam setups. 
The bands indicate the dependence on ${\theta}_{13} \in [5.7^\circ,10^\circ]$ (the lower bound is the 3$\sigma$ limit of Daya Bay). 
Left panel: results for T2HK (blue, dashed lines) and LBNE (red, solid lines). 
Right panel: results for the SPL (blue, dashed lines) and C2P (red, solid lines) setups. 
A true normal hierarchy has been assumed, and no sign degeneracies have been taken into account. 
 }
\label{fig:compDeltaRelSB}
\end{center}
\end{figure}

Although larger matter effects imply also worse precision at the optimal points, it could be interesting to combine super-beams in the
{\it vacuum} and {\it matter} regimes to reduce the dependence of $\Delta\delta$ on ${\delta}$, due to the displacement of the maxima in presence of matter effects. 
Another possibility to exploit this effect would be to combine data for two neutrino beams aimed at the same detector but peaked at different energies, 
as the proposed setup in Ref.~\cite{Bishai:2012ss}. In this case, even if the baseline is the same, the value of $\hat{A}$ would be very different,
therefore providing the desired effect.

\begin{figure}[htb!]
\begin{center}
	\includegraphics[width=15cm]{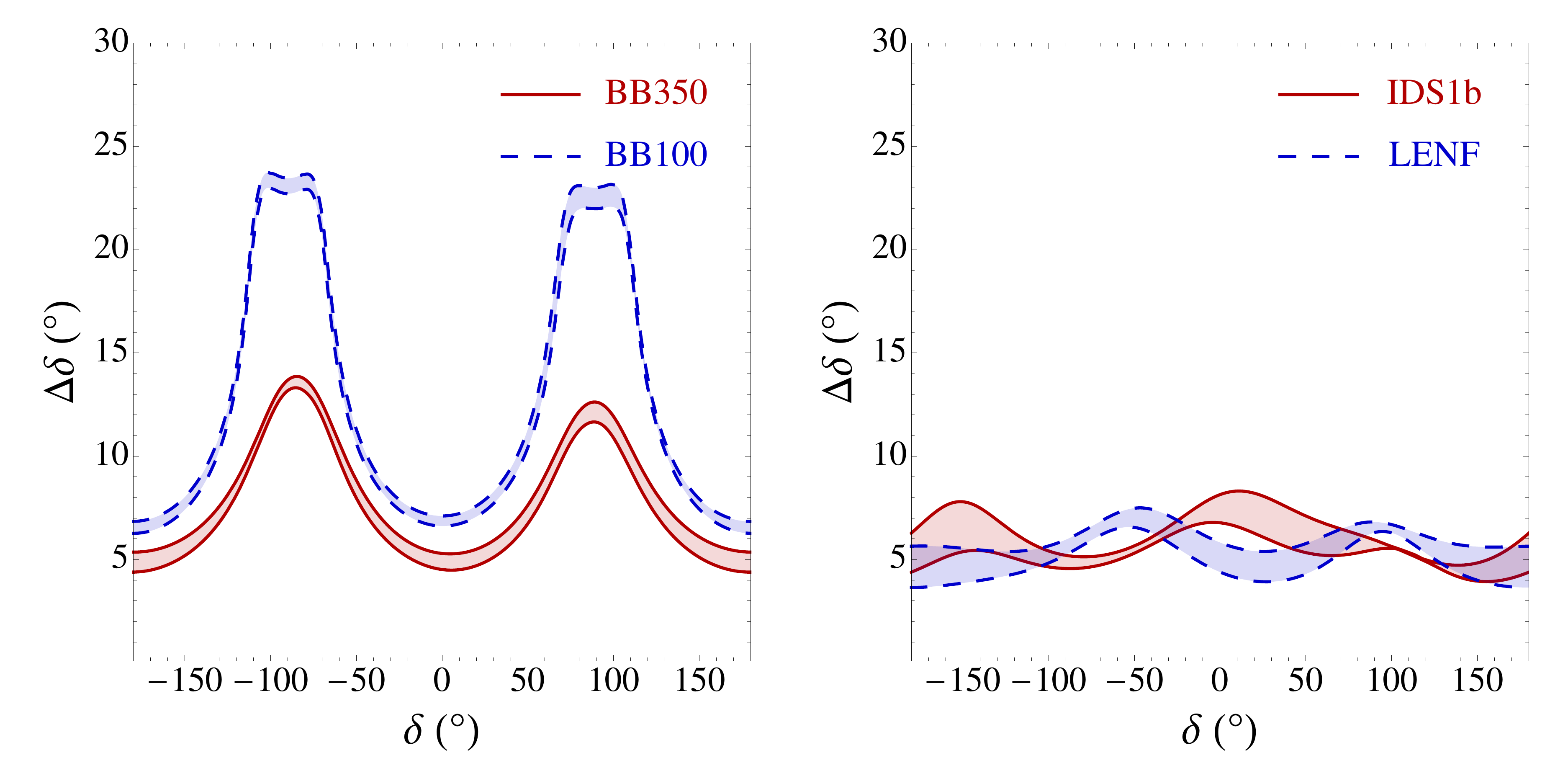}
\caption{Error on $\delta$ as a function of $ \delta$ at $1 \sigma$ (1 d.o.f.) at the considered beta-beam and neutrino factory setups. 
The bands indicate the dependence on ${\theta}_{13} \in [5.7^\circ,10^\circ]$ (the lower bound is the 3$\sigma$ limit of Daya Bay). 
Left panel: results for BB100 (blue, dashed lines) and BB350 (red, solid lines). 
Right panel: results for the LENF (blue, dashed lines) and IDS1b (red, solid lines). 
A true normal hierarchy has been assumed, and no sign degeneracies have been taken into account. 
 }
\label{fig:compDeltaRelBBNF}
\end{center}
\end{figure}

In Fig.~\ref{fig:compDeltaRelBBNF} we compare the error on $\delta$ in the beta-beam and neutrino factory setups. In this case the beta-beam setups belong to the 
{\it vacuum} category and, as a result, the precision on $\delta$ at this facilities have a strong dependence on ${\delta}$. 
The BB350 (BB100) achieves a very good precision for $\delta = 0,\pi$, with $\Delta \delta \sim 5^\circ (6^\circ)$. 
The worst precision is found, as expected,  at $\delta = \pm \pi/2$ for both setups. Their precision for maximal CP violation is, however, rather different: whereas for the BB350
we get $\Delta \delta \sim 14^\circ$ (similar to what we have found for T2HK and the SPL), the precision at the BB100 is significantly worse, $\Delta \delta \sim 23^\circ$. 
Indeed, besides the expected degradation of the measurement of $\delta$ around $\delta = \pm \pi/2$, we find that the intrinsic degeneracies are not solved in this case. This is due to the limited energy resolution 
of our simulation, and this problem may be alleviated with a more updated study of the detector response to the beta-beam fully exploiting its energy resolution capabilities.
 The neutrino factory setups belong to the {\it matter} regime (in fact, they have the strongest matter effects of all facilities) and the dependence of $\Delta \delta$
 on ${\delta}$ is, therefore, strongly suppressed. Both setups have very similar performances, with $\Delta \delta$ ranging from $4^\circ$ to $7^\circ$.   Note that the precision on $\delta$ achieved at neutrino factories and beta-beam setups 
 for $\delta \sim 0, \pi$ is also very similar. Both type of beams, therefore, are comparable in their ability to discover CP violation. On the other hand, beta-beams 
 have a worse average precision on $\delta$.
 
 We have found, however, that the performance of the neutrino factory setups is strongly affected by the assumption on the systematic error on the matter density. 
 If we increase the matter density uncertainty from $2 \%$ to $5 \%$, the precision on $\delta$ gets worse by approximately  $3^\circ$ in the whole $\delta$ range.
 The effect is much less relevant in other facilities.

As a final remark, we have checked the impact of adding a prior on $\theta_{13}$ corresponding to the expected ultimate precision of the Daya Bay experiment to each facility. 
The most remarkable effect was the resolution of the intrinsic degeneracy for the BB100 setup that improves its precision by $\sim 4^\circ$ at the worse points. 
The performance of T2HK and the SPL also improved $1^\circ-2^\circ$ while the impact of the additional prior in the other setups was rather mild.

\section{Conclusions}
\label{sec:concl}

We have studied the precision on the parameters $\theta_{13}$ and $\delta$ that would be attainable at future neutrino oscillation experiments, 
assuming that the true value of $\theta_{13}$ is in the range indicated by the recent measurements of T2K and Daya Bay. We have simulated  the performance of various setups using
different neutrino beam technologies: conventional neutrino beams and super-beams, beta-beams and neutrino factories. We have compared their performance in terms of the relative precision in $\theta_{13}$ ($\Delta_r \theta_{13}$),  and the absolute precision in $\delta$, $\Delta \delta$.  

The error on the CP phase  depends on the true values of the parameters  very significantly. As a result, measuring the performance of an experiment only 
in terms of the CP discovery potential, which is sensitive to the precision only close to the points $\delta=0, \pi$, can be misleading in some cases.
 The basic qualitative features of the dependence of $\Delta\delta$ and $\Delta_r \theta_{13}$ on the true values of the parameters can be understood from simple arguments using the approximate oscillation probabilities in the golden channel, as shown in Sec.~\ref{sec:precision}. In particular, when the baseline 
 corresponds to vacuum oscillations, the worst precision in $\delta$ corresponds precisely to the points where CP violation is maximal, $\delta=\pm \pi/2$. This is modified when matter effects are large.

\begin{figure}[htb]
\begin{center}
	\includegraphics[width=15cm]{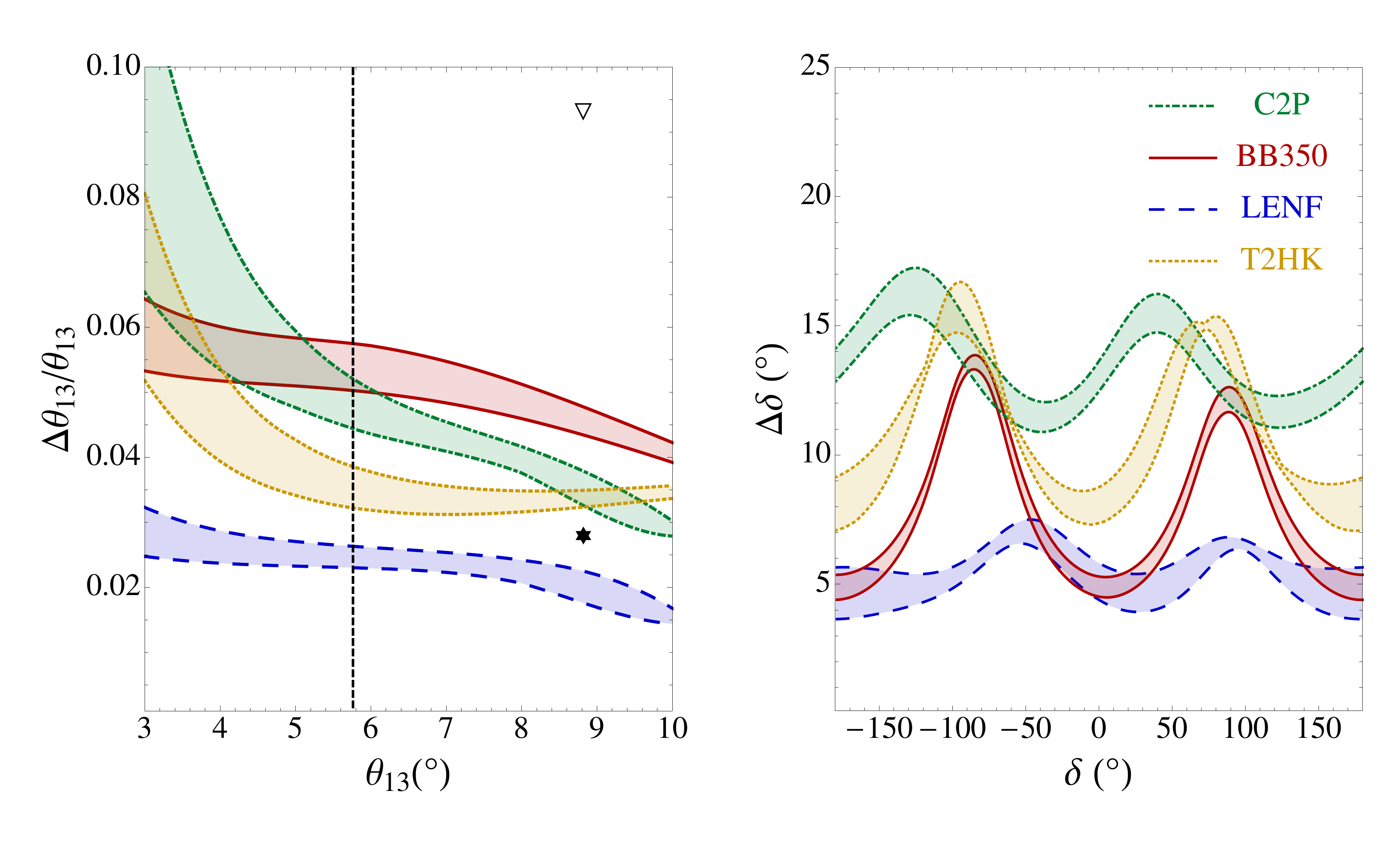}
\caption{$1\sigma$ (1 d.o.f.) precision on $\theta_{13}$ (left panel) and $\delta$ (right panel)  for the C2P  (green, dashed-dotted lines)
and T2HK (yellow, dotted lines) super-beams; the $\gamma = 350$ beta-beam (red, solid lines); and the 10 GeV Low-Energy Neutrino Factory (blue, dashed lines). 
A true normal hierarchy has been assumed, and no sign degeneracies have been taken into account. On the left panel, the empty triangle represents the present precision 
at $1\sigma$ for Daya Bay, while the star represents the ultimate attainable precision, corresponding only to the quoted systematic error.  
Both points are shown for the present best fit from Daya Bay only.  The width of the bands in each panel represent the dependence of $\Delta_r\theta_{13}$ on $\delta$ (left panel) and the dependence of $\Delta \delta$ on $\theta_{13}$ when it is varied in the range $\left[ 5.7^\circ,10^\circ\right] $ (right panel).  }
\label{fig:compBest}
\end{center}
\end{figure}

Fig.~\ref{fig:compBest} shows the comparison of four representative setups. These are: T2HK and the  CERN to Pyhas\"almi (C2P) super-beams, the $\gamma=350$ beta-beam (BB350) and the $10$ GeV low-energy neutrino factory (LENF).  It should be stressed, however, that the other super-beam and neutrino factory setups 
have similar performances.

Regarding the precision in $\theta_{13}$ and in $\delta$, neutrino factories are the optimal setup for both observables. They can  reach a 1.5\%-2.5\% accuracy in $\theta_{13}$ and measure the CP phase with an error better than $7^\circ$. The super-beams  outperform the beta-beam (but not Daya Bay) in the precision on $\theta_{13}$, but the latter can do significantly better in CP violation, except in a small region around maximal CP violation where they are comparable. 

The results for $\theta_{13}$ in the beta-beam setups are worse than in super-beams because the former  are statistically limited when compared to the SPL or T2HK, while operating at too short a baseline ({\em i.e.} with small matter effects, see Tab.~\ref{tab:summary} ) when compared to the LBNE and C2P setups. 
Regarding the precision on $\delta$,  the performance of the beta-beams around  $\delta=0,\pi$ is at the level of the neutrino factories, while it is much worse around $\delta=\pm \pi/2$. Although maximal error is expected at these points because matter effects are small, in the case of the beta-beams the deterioration is aggravated by their inability to measure the atmospheric parameters at the level of a super-beam or a neutrino factory. When the most aggressive $\gamma=350$ beta-beam is combined with the disappearance data from T2K, we find that its performance in $\Delta\delta$ is better than that of the super-beam setups considered in this paper. 

In any case, our results indicate that super-beams are clearly well fitted for the race to discover CP violation and measure $\delta$ with reasonably good precision. 
A combination of super-beams operating in the {\em vacuum} and {\em matter} regimes could reduce the large dependence of $\Delta\delta$ on the true value of the CP phase. Alternatively,  one could combine data for two neutrino beams aimed at the same detector but peaked at different energies, as the proposed setup in Ref.~\cite{Bishai:2012ss}. 

We should stress however that the performance of the facilities that we have presented depends significantly on the assumed systematic errors. If 
any of our hypothesis about fluxes, detector systematics or parameter systematics (such as the error on the matter density)  would turn out to be very different, the conclusions concerning the relative merit of each setup could change significantly. 

\section*{Acknowledgements}

We would like to thank A.~Longhin and M.~Bishai for providing the fluxes for the C2P and LBNE setups, and L.~Whitehead for providing the migration matrices for the NC backgrounds. PC would like to acknowledge useful discussions with Patrick Huber and Walter Winter, and would like to thank the support and hospitality of the Theory Division at CERN where part of this work was done. This work was partially supported by the Spanish Ministry for Education and Science projects  FPA2007-60323, FPA2009-09017, FPA2011-29678; the Consolider-Ingenio CUP (CSD2008-00037) and CPAN (CSC2007-00042); the Generalitat Valenciana (PROMETEO/2009/116); the Comunidad Aut\'onoma de Madrid (HEPHACOS P-ESP-00346 and HEPHACOS S2009/ESP-1473);  the European projects EURONU (CE212372), INVISIBLES (Marie Curie Actions, PITN-GA-2011-289442), EuCARD (European Coordination for Accelerator Research and Development, Grant Agreement number 227579) and LAGUNA (Project Number 212343). This work has been supported by the U.S. Department of Energy under award number DE-SC0003915.

\appendix
\section{Appendix}
\label{sec:appendix}

In this appendix we present the technical details of the simulations performed for the various setups included in this paper. 

\begin{table}[htb]
\begin{center}{
\renewcommand{\arraystretch}{1.3}
 \begin{tabular}{c|ccccc}
   & $L$ (km) & Det. (kton) & MW &  $(t_\nu,t_{\bar\nu})$ & Refs. \\ \hline \hline
T2K & 295 & WC (22.5) & 0.75 & (5,0) & \multirow{2}{*}{ \cite{Huber:2009cw,Ayres:2004js,Yang2004,Fechner2006,Kato}} \cr 
NO$\nu$A & 810 & TASD (15) & 0.7 & (3,3) &  \\ \hline
T2HK & 295 & WC (500) & 4.0 & (4,4) &  \cite{Huber:2002mx,Itow:2001ee,Ishitsuka:2005qi} \\ \hline
LBNE & 1290 & LAr (33.4) & 0.7 & (5,5) & \cite{Akiri:2011dv,mary,lisa} \\ \hline
SPL & 130 & WC (440) & 4.0 & (2,8) & \cite{Campagne:2006yx,Longhin:2011hn} \\ \hline 
C2P & 2300 & LAr (100) & 0.8 & (5,5) & \cite{Longhin:2010zz,Agarwalla:2011hh,Akiri:2011dv}\\ \hline\hline
 \end{tabular}}
\caption{Summary of the main details for the conventional beam and super-beam setups that have been presented in the comparison. From left to right, each column indicates the baseline, the detector technology (Water \v{C}erenkov, Totally Active Scintillator Detector, or Liquid Argon) and its fiducial mass, the beam power, the number of years that the experiment would be running in $\nu$ and $\bar\nu$ modes, and the references which have been followed in order to simulate each setup. It should be noted that the numbers quoted in this table correspond to the values stated in the cited references, where the beam power in each case has been computed according to a certain number of useful seconds per year (which in general do not coincide): T2K and T2HK assume 130 useful days per year ($1.12\times10^7$ secs, approx.);  NO$\nu$A assumes 1.7$\times 10^7$ sec$\times$yr$^{-1}$; LBNE assumes $2\times 10^7$ sec$\times$yr$^{-1}$; and SPL and C2P assume $10^7$ sec$\times$yr$^{-1}$.  \label{tab:superbeams} }
\end{center}
\end{table}

Tab.~\ref{tab:superbeams} summarizes the main features of the conventional and super-beam experiments which have been presented in the comparison. We have included the combination of T2K~\cite{Itow:2001ee} and NO$\nu$A~\cite{Ayres:2004js}, simulated as in Ref.~\cite{Huber:2009cw}. Their combination describes what can be obtained in terms of precision for $\delta$ without building any other neutrino oscillation facility. 

We have also included in our comparison the T2HK~\cite{Itow:2001ee,Nakamura:2003hk,Abe:2011ts} and LBNE proposals~\cite{Akiri:2011dv}. In the case of LBNE, the fluxes and migration matrices that have been used to simulate the response of the detector have been kindly provided by the LBNE collaboration~\cite{mary,lisa}. Fluxes in this case correspond to 120 GeV protons and $7.3\times 10^{20}$ PoT per year.

We have also considered two super-beam setups proposed in Europe, namely the SPL option as well as a setup with a much longer baseline (2300 km, from CERN to Pyhas\"almi) aiming at a LAr detector, C2P. For the SPL, the implementation of the water \v Cerenkov detector has been performed according to Ref.~\cite{Campagne:2006yx}. The flux has been provided by A.~Longhin~\cite{Longhin:2011hn} for 4.5 GeV protons, assuming 5.6$\times 10^{22}$ PoT per year. For C2P, the implementation of the LAr detector has been done according to Refs.~\cite{Akiri:2011dv,LAGUNAmeeting}. In this case, fluxes correspond to $1\times10^{21}$ PoT per year for a proton energy of 50 GeV, and have been provided by A.~Longhin~\cite{Longhin:2010zz}.

For the majority of the setups described above, systematic errors are taken as constant normalization errors over the signal and background rates. Therefore, they are correlated between different energy bins, but uncorrelated between different channels (a $5\%$ uncertainty was assumed for all of them). However, in some cases the treatment of systematics is either more sophisticated (as it is the case for T2K and NO$\nu$A) or takes into account the effect of a near detector (this is the case of LBNE). We refer the interested reader to the references quoted in Tab.~\ref{tab:superbeams}.

In the case of beta-beams, the standard ion fluxes considered in the literature are  $1.1 (2.8) \times 10^{18}$ useful \Ne ($^6$He) decays per year.  The total neutrino flux depends also on the $\gamma$ factor. In the original proposal~\cite{Zucchelli:2002sa}, ions were boosted using the existing CERN accelerator complex and, therefore, $\gamma \simeq 100$
was chosen. Due to this limitation, the neutrino flux is roughly an order of magnitude worse than that of super-beams and for this reason the physics reach of low-$\gamma$ 
beta-beams is generally limited when compared to multi-MW super-beams such as T2HK or the SPL proposals. In order to improve the statistics, higher $\gamma$ factors 
(that could be reached at CERN only with a new, refurbished, SPS) have been proposed to increase the energy of the beam. Therefore, we have included in our comparison both the original proposal as well as a variation where the boost factor is increased up to $\gamma=350$~\cite{BurguetCastell:2003vv, BurguetCastell:2005pa}. In both cases, the migration matrices and efficiencies for a WC detector exposed to a beta-beam have been taken from Ref.~\cite{BurguetCastell:2005pa}. Finally, as commented in Sec.~\ref{sec:setups}, we have observed that the precision on the atmospheric parameters has a relevant impact on the precision available for the beta-beam setups around $\delta=\pm 90^\circ$. Therefore, we have combined them with the disappearance data that can be obtained at T2K, which has been simulated according to the details in Tab.~\ref{tab:superbeams}.

A further limitation for the physics reach of beta-beams, which is particularly relevant in the case of low-$\gamma$ setups, is the atmospheric backgrounds expected at low energies (see, for instance, Ref.~\cite{FernandezMartinez:2009hb}). However, this background is most relevant for small values of $\theta_{13}$. In order to suppress the atmospheric background at the detector, the ions would be stored in very small bunches, occupying only a very small fraction of the storage ring. This is known as the suppression factor. However, the 
atmospheric background is much less troublesome for the present best fit for $\theta_{13}$ from Daya Bay and this requirement could probably be relaxed, with a consequent increase in the number of useful ion decays per year. The atmospheric background has not been included in any of the simulations presented in this paper. We have checked, though, that the inclusion of the atmospheric background with a conservative $10^{-2}$ suppression factor and a moderate $20\%$ increase in the flux actually improves slightly the results presented. 

Systematic uncertainties at a beta-beam are expected to be much smaller than at a super-beam experiment. Therefore, the constant normalization systematic errors have been set to $2.5\%$ and $5\%$ for the signal and background, respectively, for the two beta-beam setups under consideration. These are fully correlated between the different bins, but uncorrelated between different channels.

Finally, two different Neutrino Factories (NF) have also been included in our comparison. We have only included one-baseline setups, since the main purpose of placing a second detector at the magic baseline was to lift degeneracies in the case of a very small $\theta_{13}$. We have included a high energy setup, with a baseline of 4000 km and a parent muon energy of 25 GeV (which is just a modification of the setup in Ref.~\cite{NF:2011aa}, albeit with doubled neutrino flux, since all muons can be aimed at a single detector), 
and a low energy version, with a parent muon energy of 10 GeV and a baseline of 2000 km, following Ref.~\cite{Agarwalla:2010hk}. In both cases, a 100 kton MIND detector has been simulated using the migration matrices from Ref.~\cite{Laing:2010zza}. Even though these migration matrices were computed for the appearance channels, we have also used them for the disappearance channels. This may be too conservative since the cuts needed to reduce the backgrounds for the appearance signal could probably be relaxed for the disappearance signal. However, the very large statistics of the disappearance channel would largely compensate for any possible effect.

The NF is also expected to have low systematic errors. Therefore, constant normalization errors of $2.5\%$ and $5\%$ have been considered for the signal and the background, respectively, for the two setups under consideration. These are fully correlated between different bins, but uncorrelated between different channels.


\begin{thebibliography}{10}

\bibitem{An:2012eh}
{\bfseries DAYA-BAY} Collaboration, F.~P. An {\em et al.}, ``{Observation of
  electron-antineutrino disappearance at Daya Bay},''
\href{http://arxiv.org/abs/1203.1669}{{\ttfamily arXiv:1203.1669 [hep-ex]}}.

\bibitem{Abe:2011sj}
{\bfseries T2K Collaboration} Collaboration, K.~Abe {\em et al.}, ``{Indication
  of Electron Neutrino Appearance from an Accelerator-produced Off-axis Muon
  Neutrino Beam},'' {\em Phys.Rev.Lett.} {\bfseries 107} (2011) 041801,
  \href{http://arxiv.org/abs/1106.2822}{{\ttfamily arXiv:1106.2822 [hep-ex]}}.

\bibitem{Abe:2011fz}
{\bfseries DOUBLE-CHOOZ} Collaboration, Y.~Abe {\em et al.}, ``{Indication for
  the disappearance of reactor electron antineutrinos in the Double Chooz
  experiment},''
\href{http://arxiv.org/abs/1112.6353}{{\ttfamily arXiv:1112.6353 [hep-ex]}}.

\bibitem{Adamson:2011qu}
{\bfseries MINOS Collaboration} Collaboration, P.~Adamson {\em et al.},
  ``{Improved search for muon-neutrino to electron-neutrino oscillations in
  MINOS},'' \href{http://dx.doi.org/10.1103/PhysRevLett.107.181802}{{\em
  Phys.Rev.Lett.} {\bfseries 107} (2011) 181802},
\href{http://arxiv.org/abs/1108.0015}{{\ttfamily arXiv:1108.0015 [hep-ex]}}.

\bibitem{GonzalezGarcia:2010er}
M.~Gonzalez-Garcia, M.~Maltoni, and J.~Salvado, ``{Updated global fit to three
  neutrino mixing: status of the hints of theta13 $>$ 0},''
  \href{http://dx.doi.org/10.1007/JHEP04(2010)056}{{\em JHEP} {\bfseries 1004}
  (2010) 056},
\href{http://arxiv.org/abs/1001.4524}{{\ttfamily arXiv:1001.4524 [hep-ph]}}.

\bibitem{Fogli:2011qn}
G.~Fogli, E.~Lisi, A.~Marrone, A.~Palazzo, and A.~Rotunno, ``{Evidence of
  $\theta_{13}>0$ from global neutrino data analysis},''
  \href{http://dx.doi.org/10.1103/PhysRevD.84.053007}{{\em Phys.Rev.}
  {\bfseries D84} (2011) 053007},
  \href{http://arxiv.org/abs/1106.6028}{{\ttfamily arXiv:1106.6028 [hep-ph]}}.

\bibitem{Fukugita:1986hr}
M.~Fukugita and T.~Yanagida, ``{Baryogenesis Without Grand Unification},''
\href{http://dx.doi.org/10.1016/0370-2693(86)91126-3}{{\em Phys. Lett.}
  {\bfseries B174} (1986) 45}.

\bibitem{Bandyopadhyay:2007kx}
{\bfseries ISS Physics Working Group} Collaboration, A.~Bandyopadhyay {\em et
  al.}, ``{Physics at a future Neutrino Factory and super-beam facility},''
  \href{http://dx.doi.org/10.1088/0034-4885/72/10/106201}{{\em Rept. Prog.
  Phys.} {\bfseries 72} (2009) 106201},
\href{http://arxiv.org/abs/0710.4947}{{\ttfamily arXiv:0710.4947 [hep-ph]}}.

\bibitem{Geer:1997iz}
S.~Geer, ``{Neutrino beams from muon storage rings: Characteristics and physics
  potential},'' \href{http://dx.doi.org/10.1103/PhysRevD.57.6989,
  10.1103/PhysRevD.59.039903}{{\em Phys.Rev.} {\bfseries D57} (1998)
  6989--6997},
\href{http://arxiv.org/abs/hep-ph/9712290}{{\ttfamily arXiv:hep-ph/9712290
  [hep-ph]}}.

\bibitem{DeRujula:1998hd}
A.~De~Rujula, M.~Gavela, and P.~Hernandez, ``{Neutrino oscillation physics with
  a neutrino factory},''
  \href{http://dx.doi.org/10.1016/S0550-3213(99)00070-X}{{\em Nucl.Phys.}
  {\bfseries B547} (1999) 21--38},
  \href{http://arxiv.org/abs/hep-ph/9811390}{{\ttfamily arXiv:hep-ph/9811390
  [hep-ph]}}.

\bibitem{Apollonio:2002en}
M.~Apollonio, A.~Blondel, A.~Broncano, M.~Bonesini, J.~Bouchez, {\em et al.},
  ``{Oscillation physics with a neutrino factory},''
  \href{http://arxiv.org/abs/hep-ph/0210192}{{\ttfamily arXiv:hep-ph/0210192
  [hep-ph]}}.
To appear on the CERN Yellow Report on the Neutrino Factory.

\bibitem{GomezCadenas:2002fz}
J.~Gomez-Cadenas and D.~A. Harris, ``{Physics opportunities at neutrino
  factories},''
\href{http://dx.doi.org/10.1146/annurev.nucl.52.050102.090653}{{\em
  Ann.Rev.Nucl.Part.Sci.} {\bfseries 52} (2002) 253--302}.

\bibitem{Zucchelli:2002sa}
P.~Zucchelli, ``{A novel concept for a anti-nu/e / nu/e neutrino factory: The
  beta beam},'' \href{http://dx.doi.org/10.1016/S0370-2693(02)01576-9}{{\em
  Phys.Lett.} {\bfseries B532} (2002) 166--172}.

\bibitem{NF:2011aa}
{\bfseries IDS-NF Collaboration} Collaboration, S.~Choubey {\em et al.},
  ``{International Design Study for the Neutrino Factory, Interim Design
  Report},''
\href{http://arxiv.org/abs/1112.2853}{{\ttfamily arXiv:1112.2853 [hep-ex]}}.

\bibitem{Huber:2005ep}
P.~Huber, M.~Maltoni, and T.~Schwetz, ``{Resolving parameter degeneracies in
  long-baseline experiments by atmospheric neutrino data},''
  \href{http://dx.doi.org/10.1103/PhysRevD.71.053006}{{\em Phys.Rev.}
  {\bfseries D71} (2005) 053006},
\href{http://arxiv.org/abs/hep-ph/0501037}{{\ttfamily arXiv:hep-ph/0501037
  [hep-ph]}}.

\bibitem{Campagne:2006yx}
J.-E. Campagne, M.~Maltoni, M.~Mezzetto, and T.~Schwetz, ``{Physics potential
  of the CERN-MEMPHYS neutrino oscillation project},''
  \href{http://dx.doi.org/10.1088/1126-6708/2007/04/003}{{\em JHEP} {\bfseries
  0704} (2007) 003}, \href{http://arxiv.org/abs/hep-ph/0603172}{{\ttfamily
  arXiv:hep-ph/0603172 [hep-ph]}}.

\bibitem{TabarellideFatis:2002ni}
T.~Tabarelli~de Fatis, ``{Prospects of measuring sin**2 2 Theta(13) and the
  sign of Delta m**2 with a massive magnetized detector for atmospheric
  neutrinos},'' \href{http://dx.doi.org/10.1007/s100520200935}{{\em
  Eur.Phys.J.} {\bfseries C24} (2002) 43--50},
  \href{http://arxiv.org/abs/hep-ph/0202232}{{\ttfamily arXiv:hep-ph/0202232
  [hep-ph]}}.
(8 pages, 8 figures, submitted to Eur.Phys.J.C) Report-no: Bicocca-EX-02-01.

\bibitem{Bernabeu:2003yp}
J.~Bernabeu, S.~Palomares~Ruiz, and S.~Petcov, ``{Atmospheric neutrino
  oscillations, theta(13) and neutrino mass hierarchy},''
  \href{http://dx.doi.org/10.1016/j.nuclphysb.2003.07.025}{{\em Nucl.Phys.}
  {\bfseries B669} (2003) 255--276},
\href{http://arxiv.org/abs/hep-ph/0305152}{{\ttfamily arXiv:hep-ph/0305152
  [hep-ph]}}.

\bibitem{PalomaresRuiz:2004tk}
S.~Palomares-Ruiz and S.~Petcov, ``{Three-neutrino oscillations of atmospheric
  neutrinos, theta(13), neutrino mass hierarchy and iron magnetized
  detectors},'' \href{http://dx.doi.org/10.1016/j.nuclphysb.2005.01.045}{{\em
  Nucl.Phys.} {\bfseries B712} (2005) 392--410},
\href{http://arxiv.org/abs/hep-ph/0406096}{{\ttfamily arXiv:hep-ph/0406096
  [hep-ph]}}.

\bibitem{Indumathi:2004kd}
D.~Indumathi and M.~Murthy, ``{A Question of hierarchy: Matter effects with
  atmospheric neutrinos and anti-neutrinos},''
  \href{http://dx.doi.org/10.1103/PhysRevD.71.013001}{{\em Phys.Rev.}
  {\bfseries D71} (2005) 013001},
\href{http://arxiv.org/abs/hep-ph/0407336}{{\ttfamily arXiv:hep-ph/0407336
  [hep-ph]}}.

\bibitem{Petcov:2005rv}
S.~Petcov and T.~Schwetz, ``{Determining the neutrino mass hierarchy with
  atmospheric neutrinos},''
  \href{http://dx.doi.org/10.1016/j.nuclphysb.2006.01.020}{{\em Nucl.Phys.}
  {\bfseries B740} (2006) 1--22},
\href{http://arxiv.org/abs/hep-ph/0511277}{{\ttfamily arXiv:hep-ph/0511277
  [hep-ph]}}.

\bibitem{Arumugam:2005nt}
{\bfseries INO Collaboration} Collaboration, V.~Arumugam {\em et al.},
``{India-based Neutrino Observatory: Interim project report. Vol. 1},''.

\bibitem{Samanta:2006sj}
A.~Samanta, ``{The Mass hierarchy with atmospheric neutrinos at INO},''
  \href{http://dx.doi.org/10.1016/j.physletb.2009.01.067}{{\em Phys.Lett.}
  {\bfseries B673} (2009) 37--46},
\href{http://arxiv.org/abs/hep-ph/0610196}{{\ttfamily arXiv:hep-ph/0610196
  [hep-ph]}}.

\bibitem{Gandhi:2007td}
R.~Gandhi, P.~Ghoshal, S.~Goswami, P.~Mehta, S.~Sankar, {\em et al.}, ``{Mass
  Hierarchy Determination via future Atmospheric Neutrino Detectors},''
  \href{http://dx.doi.org/10.1103/PhysRevD.76.073012}{{\em Phys.Rev.}
  {\bfseries D76} (2007) 073012},
\href{http://arxiv.org/abs/0707.1723}{{\ttfamily arXiv:0707.1723 [hep-ph]}}.

\bibitem{Mena:2008rh}
O.~Mena, I.~Mocioiu, and S.~Razzaque, ``{Neutrino mass hierarchy extraction
  using atmospheric neutrinos in ice},''
  \href{http://dx.doi.org/10.1103/PhysRevD.78.093003}{{\em Phys.Rev.}
  {\bfseries D78} (2008) 093003},
\href{http://arxiv.org/abs/0803.3044}{{\ttfamily arXiv:0803.3044 [hep-ph]}}.

\bibitem{Gandhi:2008zs}
R.~Gandhi, P.~Ghoshal, S.~Goswami, and S.~Sankar, ``{Resolving the Mass
  Hierarchy with Atmospheric Neutrinos using a Liquid Argon Detector},''
  \href{http://dx.doi.org/10.1103/PhysRevD.78.073001}{{\em Phys.Rev.}
  {\bfseries D78} (2008) 073001},
  \href{http://arxiv.org/abs/0807.2759}{{\ttfamily arXiv:0807.2759 [hep-ph]}}.
13 pages, 3 figures.

\bibitem{Samanta:2009qw}
A.~Samanta, ``{Discrimination of mass hierarchy with atmospheric neutrinos at a
  magnetized muon detector},''
  \href{http://dx.doi.org/10.1103/PhysRevD.81.037302}{{\em Phys.Rev.}
  {\bfseries D81} (2010) 037302},
\href{http://arxiv.org/abs/0907.3540}{{\ttfamily arXiv:0907.3540 [hep-ph]}}.

\bibitem{Blennow:2012gj}
M.~Blennow and T.~Schwetz, ``{Identifying the Neutrino mass Ordering with INO
  and NOvA},''
\href{http://arxiv.org/abs/1203.3388}{{\ttfamily arXiv:1203.3388 [hep-ph]}}.

\bibitem{Winter:2003ye}
W.~Winter, ``{Understanding CP phase dependent measurements at neutrino
  superbeams in terms of bi-rate graphs},''
  \href{http://dx.doi.org/10.1103/PhysRevD.70.033006}{{\em Phys.Rev.}
  {\bfseries D70} (2004) 033006},
\href{http://arxiv.org/abs/hep-ph/0310307}{{\ttfamily arXiv:hep-ph/0310307
  [hep-ph]}}.

\bibitem{Huber:2004gg}
P.~Huber, M.~Lindner, and W.~Winter, ``{From parameter space constraints to the
  precision determination of the leptonic Dirac CP phase},''
  \href{http://dx.doi.org/10.1088/1126-6708/2005/05/020}{{\em JHEP} {\bfseries
  0505} (2005) 020},
\href{http://arxiv.org/abs/hep-ph/0412199}{{\ttfamily arXiv:hep-ph/0412199
  [hep-ph]}}.


\bibitem{Huber:2004ka}
P.~Huber, M.~Lindner, and W.~Winter, ``{Simulation of long-baseline neutrino
  oscillation experiments with GLoBES (General Long Baseline Experiment
  Simulator)},'' \href{http://dx.doi.org/10.1016/j.cpc.2005.01.003}{{\em
  Comput.Phys.Commun.} {\bfseries 167} (2005) 195},
  \href{http://arxiv.org/abs/hep-ph/0407333}{{\ttfamily arXiv:hep-ph/0407333
  [hep-ph]}}.

\bibitem{Huber:2007ji}
P.~Huber, J.~Kopp, M.~Lindner, M.~Rolinec, and W.~Winter, ``{New features in
  the simulation of neutrino oscillation experiments with GLoBES 3.0: General
  Long Baseline Experiment Simulator},''
  \href{http://dx.doi.org/10.1016/j.cpc.2007.05.004}{{\em Comput.Phys.Commun.}
  {\bfseries 177} (2007) 432--438},
  \href{http://arxiv.org/abs/hep-ph/0701187}{{\ttfamily arXiv:hep-ph/0701187
  [hep-ph]}}.

\bibitem{Akiri:2011dv}
{\bfseries LBNE Collaboration} Collaboration, T.~Akiri {\em et al.}, ``{The
  2010 Interim Report of the Long-Baseline Neutrino Experiment Collaboration
  Physics Working Groups},'' \href{http://arxiv.org/abs/1110.6249}{{\ttfamily
  arXiv:1110.6249 [hep-ex]}}.
Corresponding author R.J.Wilson (Bob.Wilson@colostate.edu)/ 113 pages, 90
  figures.

\bibitem{GomezCadenas:2001eua}
{\bfseries CERN working group on Super Beams} Collaboration, J.~J.
  Gomez-Cadenas {\em et al.}, ``{Physics potential of very intense conventional
  neutrino beams},'' \href{http://arxiv.org/abs/hep-ph/0105297}{{\ttfamily
  arXiv:hep-ph/0105297 [hep-ph]}}.
Talk given at the Venice Conference on Neutrino Telescopes, Venice, March, 2001
  Report-no: IFIC/01-31.

\bibitem{Campagne:2004wt}
J.~E. Campagne and A.~Cazes, ``{The theta(13) and delta(CP) sensitivities of
  the SPL-Frejus project revisited},''
  \href{http://dx.doi.org/10.1140/epjc/s2005-02455-x}{{\em Eur.Phys.J.}
  {\bfseries C45} (2006) 643--657},
\href{http://arxiv.org/abs/hep-ex/0411062}{{\ttfamily arXiv:hep-ex/0411062
  [hep-ex]}}.

\bibitem{Longhin:2011hn}
A.~Longhin, ``{A new design for the CERN-Fr\'ejus neutrino Super Beam},''
  \href{http://dx.doi.org/10.1140/epjc/s10052-011-1745-8}{{\em Eur.Phys.J.}
  {\bfseries C71} (2011) 1745},
  \href{http://arxiv.org/abs/1106.1096}{{\ttfamily arXiv:1106.1096
  [physics.acc-ph]}}.

\bibitem{Angus:2010sz}
{\bfseries LAGUNA Collaboration} Collaboration, D.~Angus {\em et al.}, ``{The
  LAGUNA design study- towards giant liquid based underground detectors for
  neutrino physics and astrophysics and proton decay searches},''
  \href{http://arxiv.org/abs/1001.0077}{{\ttfamily arXiv:1001.0077
  [physics.ins-det]}}.

\bibitem{Itow:2001ee}
{\bfseries The T2K Collaboration} Collaboration, Y.~Itow {\em et al.}, ``{The
  JHF-Kamioka neutrino project},''
  \href{http://arxiv.org/abs/hep-ex/0106019}{{\ttfamily arXiv:hep-ex/0106019
  [hep-ex]}}.

\bibitem{Nakamura:2003hk}
K.~Nakamura, ``{Hyper-Kamiokande: A next generation water Cherenkov
  detector},'' \href{http://dx.doi.org/10.1142/S0217751X03017361}{{\em
  Int.J.Mod.Phys.} {\bfseries A18} (2003) 4053--4063}.

\bibitem{Abe:2011ts}
K.~Abe, T.~Abe, H.~Aihara, Y.~Fukuda, Y.~Hayato, {\em et al.}, ``{Letter of
  Intent: The Hyper-Kamiokande Experiment --- Detector Design and Physics
  Potential ---},''
\href{http://arxiv.org/abs/1109.3262}{{\ttfamily arXiv:1109.3262 [hep-ex]}}.

\bibitem{Donini:2004iv}
A.~Donini, E.~Fernandez-Martinez, and S.~Rigolin, ``{Appearance and
  disappearance signals at a beta-beam and a super-beam facility},''
  \href{http://dx.doi.org/10.1016/j.physletb.2005.06.072}{{\em Phys.Lett.}
  {\bfseries B621} (2005) 276--287},
  \href{http://arxiv.org/abs/hep-ph/0411402}{{\ttfamily arXiv:hep-ph/0411402
  [hep-ph]}}.

\bibitem{Donini:2005rn}
A.~Donini, D.~Meloni, and S.~Rigolin, ``{The Impact of solar and atmospheric
  parameter uncertainties on the measurement of theta(13) and delta},''
  \href{http://dx.doi.org/10.1140/epjc/s2005-02416-5}{{\em Eur.Phys.J.}
  {\bfseries C45} (2006) 73--95},
\href{http://arxiv.org/abs/hep-ph/0506100}{{\ttfamily arXiv:hep-ph/0506100
  [hep-ph]}}.

\bibitem{BurguetCastell:2003vv}
J.~Burguet-Castell, D.~Casper, J.~Gomez-Cadenas, P.~Hernandez, and F.~Sanchez,
  ``{Neutrino oscillation physics with a higher gamma beta beam},''
  \href{http://dx.doi.org/10.1016/j.nuclphysb.2004.06.021}{{\em Nucl.Phys.}
  {\bfseries B695} (2004) 217--240},
  \href{http://arxiv.org/abs/hep-ph/0312068}{{\ttfamily arXiv:hep-ph/0312068
  [hep-ph]}}.

\bibitem{BurguetCastell:2005pa}
J.~Burguet-Castell, D.~Casper, E.~Couce, J.~Gomez-Cadenas, and P.~Hernandez,
  ``{Optimal beta-beam at the CERN-SPS},''
  \href{http://dx.doi.org/10.1016/j.nuclphysb.2005.06.037}{{\em Nucl.Phys.}
  {\bfseries B725} (2005) 306--326},
  \href{http://arxiv.org/abs/hep-ph/0503021}{{\ttfamily arXiv:hep-ph/0503021
  [hep-ph]}}.

\bibitem{Mezzetto:2003ub}
M.~Mezzetto, ``{Physics reach of the beta beam},''
  \href{http://dx.doi.org/10.1088/0954-3899/29/8/346}{{\em J.Phys.G} {\bfseries
  G29} (2003) 1771--1776},
  \href{http://arxiv.org/abs/hep-ex/0302007}{{\ttfamily arXiv:hep-ex/0302007
  [hep-ex]}}.
  
\bibitem{Donini:2004hu}
A.~Donini, E.~Fernandez-Martinez, P.~Migliozzi, S.~Rigolin, and
  L.~Scotto~Lavina, ``{Study of the eightfold degeneracy with a standard
  beta-beam and a super-beam facility},''
  \href{http://dx.doi.org/10.1016/j.nuclphysb.2004.12.029}{{\em Nucl.Phys.}
  {\bfseries B710} (2005) 402--424},
  \href{http://arxiv.org/abs/hep-ph/0406132}{{\ttfamily arXiv:hep-ph/0406132
  [hep-ph]}}.

\bibitem{Mezzetto:2004gs}
M.~Mezzetto, ``{Beta beams},''
  \href{http://dx.doi.org/10.1016/j.nuclphysbps.2005.01.123}{{\em
  Nucl.Phys.Proc.Suppl.} {\bfseries 143} (2005) 309--316},
  \href{http://arxiv.org/abs/hep-ex/0410083}{{\ttfamily arXiv:hep-ex/0410083
  [hep-ex]}}.

\bibitem{Huber:2005jk}
P.~Huber, M.~Lindner, M.~Rolinec, and W.~Winter, ``{Physics and optimization of
  beta-beams: From low to very high gamma},''
  \href{http://dx.doi.org/10.1103/PhysRevD.73.053002}{{\em Phys.Rev.}
  {\bfseries D73} (2006) 053002},
  \href{http://arxiv.org/abs/hep-ph/0506237}{{\ttfamily arXiv:hep-ph/0506237
  [hep-ph]}}.

\bibitem{FernandezMartinez:2009hb}
E.~Fernandez-Martinez, ``{The gamma = 100 beta-Beam revisited},''
  \href{http://dx.doi.org/10.1016/j.nuclphysb.2010.02.028}{{\em Nucl.Phys.}
  {\bfseries B833} (2010) 96--107},
  \href{http://arxiv.org/abs/0912.3804}{{\ttfamily arXiv:0912.3804 [hep-ph]}}.

\bibitem{Agarwalla:2005we}
S.~K. Agarwalla, A.~Raychaudhuri, and A.~Samanta, ``{Exploration prospects of a
  long baseline beta beam neutrino experiment with an iron calorimeter
  detector},'' \href{http://dx.doi.org/10.1016/j.physletb.2005.09.053}{{\em
  Phys.Lett.} {\bfseries B629} (2005) 33--40},
  \href{http://arxiv.org/abs/hep-ph/0505015}{{\ttfamily arXiv:hep-ph/0505015
  [hep-ph]}}.

\bibitem{Donini:2006tt}
A.~Donini, E.~Fernandez-Martinez, P.~Migliozzi, S.~Rigolin, L.~Scotto~Lavina,
  {\em et al.}, ``{A Beta Beam complex based on the machine upgrades of the
  LHC},'' \href{http://dx.doi.org/10.1140/epjc/s10052-006-0019-3}{{\em
  Eur.Phys.J.} {\bfseries C48} (2006) 787--796},
  \href{http://arxiv.org/abs/hep-ph/0604229}{{\ttfamily arXiv:hep-ph/0604229
  [hep-ph]}}.

\bibitem{Volpe:2006in}
C.~Volpe, ``{Topical Review on Beta-beams},''
  \href{http://dx.doi.org/10.1088/0954-3899/34/1/R01}{{\em J.Phys.G} {\bfseries
  G34} (2007) R1--R44}, \href{http://arxiv.org/abs/hep-ph/0605033}{{\ttfamily
  arXiv:hep-ph/0605033 [hep-ph]}}.

\bibitem{Agarwalla:2006vf}
S.~K. Agarwalla, S.~Choubey, and A.~Raychaudhuri, ``{Neutrino mass hierarchy
  and theta(13) with a magic baseline beta-beam experiment},''
  \href{http://dx.doi.org/10.1016/j.nuclphysb.2007.02.012}{{\em Nucl.Phys.}
  {\bfseries B771} (2007) 1--27},
  \href{http://arxiv.org/abs/hep-ph/0610333}{{\ttfamily arXiv:hep-ph/0610333
  [hep-ph]}}.

\bibitem{Donini:2007qt}
A.~Donini, E.~Fernandez-Martinez, P.~Migliozzi, S.~Rigolin, L.~Lavina, {\em et
  al.}, ``{Neutrino hierarchy from CP-blind observables with high density
  magnetized detectors},''
  \href{http://dx.doi.org/10.1140/epjc/s10052-007-0489-y}{{\em Eur.Phys.J.}
  {\bfseries C53} (2008) 599--606},
  \href{http://arxiv.org/abs/hep-ph/0703209}{{\ttfamily arXiv:hep-ph/0703209
  [HEP-PH]}}.

\bibitem{Jansson:2007nm}
A.~Jansson, O.~Mena, S.~J. Parke, and N.~Saoulidou, ``{Combining CPT-conjugate
  neutrino channels at Fermilab},''
  \href{http://dx.doi.org/10.1103/PhysRevD.78.053002}{{\em Phys.Rev.}
  {\bfseries D78} (2008) 053002},
  \href{http://arxiv.org/abs/0711.1075}{{\ttfamily arXiv:0711.1075 [hep-ph]}}.

\bibitem{Agarwalla:2007ai}
S.~K. Agarwalla, S.~Choubey, and A.~Raychaudhuri, ``{Unraveling neutrino
  parameters with a magical beta-beam experiment at INO},''
  \href{http://dx.doi.org/10.1016/j.nuclphysb.2008.01.031}{{\em Nucl.Phys.}
  {\bfseries B798} (2008) 124--145},
  \href{http://arxiv.org/abs/0711.1459}{{\ttfamily arXiv:0711.1459 [hep-ph]}}.

\bibitem{Coloma:2007nn}
P.~Coloma, A.~Donini, E.~Fernandez-Martinez, and J.~Lopez-Pavon, ``{theta(13),
  delta and the neutrino mass hierarchy at a gamma = 350 double baseline Li/B
  beta-Beam},'' \href{http://dx.doi.org/10.1088/1126-6708/2008/05/050}{{\em
  JHEP} {\bfseries 0805} (2008) 050},
  \href{http://arxiv.org/abs/0712.0796}{{\ttfamily arXiv:0712.0796 [hep-ph]}}.

\bibitem{Meloni:2008it}
D.~Meloni, O.~Mena, C.~Orme, S.~Palomares-Ruiz, and S.~Pascoli, ``{An
  Intermediate gamma beta-beam neutrino experiment with long baseline},''
  \href{http://dx.doi.org/10.1088/1126-6708/2008/07/115}{{\em JHEP} {\bfseries
  0807} (2008) 115}, \href{http://arxiv.org/abs/0802.0255}{{\ttfamily
  arXiv:0802.0255 [hep-ph]}}.

\bibitem{Agarwalla:2008gf}
S.~K. Agarwalla, S.~Choubey, A.~Raychaudhuri, and W.~Winter, ``{Optimizing the
  greenfield Beta-beam},''
  \href{http://dx.doi.org/10.1088/1126-6708/2008/06/090}{{\em JHEP} {\bfseries
  0806} (2008) 090}, \href{http://arxiv.org/abs/0802.3621}{{\ttfamily
  arXiv:0802.3621 [hep-ex]}}.

\bibitem{Agarwalla:2008ti}
S.~K. Agarwalla, S.~Choubey, and A.~Raychaudhuri, ``{Exceptional Sensitivity to
  Neutrino Parameters with a Two Baseline Beta-Beam Set-up},''
  \href{http://dx.doi.org/10.1016/j.nuclphysb.2008.07.026}{{\em Nucl.Phys.}
  {\bfseries B805} (2008) 305--325},
  \href{http://arxiv.org/abs/0804.3007}{{\ttfamily arXiv:0804.3007 [hep-ph]}}.

\bibitem{Winter:2008cn}
W.~Winter, ``{Minimal Neutrino Beta Beam for Large theta(13)},''
  \href{http://dx.doi.org/10.1103/PhysRevD.78.037101}{{\em Phys.Rev.}
  {\bfseries D78} (2008) 037101},
  \href{http://arxiv.org/abs/0804.4000}{{\ttfamily arXiv:0804.4000 [hep-ph]}}.

\bibitem{Winter:2008dj}
W.~Winter, ``{Optimizing the green-field beta beam: Small versus large
  theta(13)},'' {\em PoS} {\bfseries NUFACT08} (2008) 036,
  \href{http://arxiv.org/abs/0809.3890}{{\ttfamily arXiv:0809.3890 [hep-ph]}}.

\bibitem{Choubey:2009ks}
S.~Choubey, P.~Coloma, A.~Donini, and E.~Fernandez-Martinez, ``{Optimized
  Two-Baseline Beta-Beam Experiment},''
  \href{http://dx.doi.org/10.1088/1126-6708/2009/12/020}{{\em JHEP} {\bfseries
  0912} (2009) 020}, \href{http://arxiv.org/abs/0907.2379}{{\ttfamily
  arXiv:0907.2379 [hep-ph]}}.

\bibitem{Coloma:2010wa}
P.~Coloma, A.~Donini, P.~Migliozzi, L.~Scotto~Lavina, and F.~Terranova, ``{A
  minimal Beta Beam with high-Q ions to address CP violation in the leptonic
  sector},'' \href{http://dx.doi.org/10.1140/epjc/s10052-011-1674-6}{{\em
  Eur.Phys.J.} {\bfseries C71} (2011) 1674},
  \href{http://arxiv.org/abs/1004.3773}{{\ttfamily arXiv:1004.3773 [hep-ph]}}.

\bibitem{BurguetCastell:2001ez}
J.~Burguet-Castell, M.~B. Gavela, J.~J. Gomez-Cadenas, P.~Hernandez, and
  O.~Mena, ``{On the measurement of leptonic CP violation},''
  \href{http://dx.doi.org/10.1016/S0550-3213(01)00248-6}{{\em Nucl. Phys.}
  {\bfseries B608} (2001) 301--318},
\href{http://arxiv.org/abs/hep-ph/0103258}{{\ttfamily arXiv:hep-ph/0103258}}.

\bibitem{Huber:2003ak}
P.~Huber and W.~Winter, ``{Neutrino factories and the 'magic' baseline},''
  \href{http://dx.doi.org/10.1103/PhysRevD.68.037301}{{\em Phys.Rev.}
  {\bfseries D68} (2003) 037301},
\href{http://arxiv.org/abs/hep-ph/0301257}{{\ttfamily arXiv:hep-ph/0301257
  [hep-ph]}}.

\bibitem{Geer:2007kn}
S.~Geer, O.~Mena, and S.~Pascoli, ``{A Low energy neutrino factory for large
  $\theta_{13}$},'' \href{http://dx.doi.org/10.1103/PhysRevD.75.093001}{{\em
  Phys.Rev.} {\bfseries D75} (2007) 093001},
  \href{http://arxiv.org/abs/hep-ph/0701258}{{\ttfamily arXiv:hep-ph/0701258
  [HEP-PH]}}.

\bibitem{Bross:2007ts}
A.~D. Bross, M.~Ellis, S.~Geer, O.~Mena, and S.~Pascoli, ``{A Neutrino factory
  for both large and small $\theta_{13}$},''
  \href{http://dx.doi.org/10.1103/PhysRevD.77.093012}{{\em Phys.Rev.}
  {\bfseries D77} (2008) 093012},
  \href{http://arxiv.org/abs/0709.3889}{{\ttfamily arXiv:0709.3889 [hep-ph]}}.

\bibitem{FernandezMartinez:2010zza}
E.~Fernandez~Martinez, T.~Li, S.~Pascoli, and O.~Mena, ``{Improvement of the
  low energy neutrino factory},''
  \href{http://dx.doi.org/10.1103/PhysRevD.81.073010}{{\em Phys.Rev.}
  {\bfseries D81} (2010) 073010},
  \href{http://arxiv.org/abs/0911.3776}{{\ttfamily arXiv:0911.3776 [hep-ph]}}.

\bibitem{Agarwalla:2010hk}
S.~K. Agarwalla, P.~Huber, J.~Tang, and W.~Winter, ``{Optimization of the
  Neutrino Factory, revisited},''
  \href{http://dx.doi.org/10.1007/JHEP01(2011)120}{{\em JHEP} {\bfseries 1101}
  (2011) 120}, \href{http://arxiv.org/abs/1012.1872}{{\ttfamily arXiv:1012.1872
  [hep-ph]}}.

\bibitem{Asano:2011nj}
K.~Asano and H.~Minakata, ``{Large-Theta(13) Perturbation Theory of Neutrino
  Oscillation for Long-Baseline Experiments},''
  \href{http://dx.doi.org/10.1007/JHEP06(2011)022}{{\em JHEP} {\bfseries 1106}
  (2011) 022},
\href{http://arxiv.org/abs/1103.4387}{{\ttfamily arXiv:1103.4387 [hep-ph]}}.

\bibitem{Cervera:2000kp}
A.~Cervera, A.~Donini, M.~Gavela, J.~Gomez~Cadenas, P.~Hernandez, {\em et al.},
  ``{Golden measurements at a neutrino factory},''
  \href{http://dx.doi.org/10.1016/S0550-3213(00)00221-2,
  10.1016/S0550-3213(00)00221-2}{{\em Nucl.Phys.} {\bfseries B579} (2000)
  17--55}, \href{http://arxiv.org/abs/hep-ph/0002108}{{\ttfamily
  arXiv:hep-ph/0002108 [hep-ph]}}.

\bibitem{Freund:2001ui}
  M.~Freund, P.~Huber and M.~Lindner,
  ``Systematic exploration of the neutrino factory parameter space  including
  errors and correlations,''
  Nucl.\ Phys.\  B {\bf 615} (2001) 331
  [arXiv:hep-ph/0105071].

\bibitem{Akhmedov:2004ny}
E.~K. Akhmedov, R.~Johansson, M.~Lindner, T.~Ohlsson, and T.~Schwetz, ``{Series
  expansions for three-flavor neutrino oscillation probabilities in matter},''
  \href{http://dx.doi.org/10.1088/1126-6708/2004/04/078}{{\em JHEP} {\bfseries
  04} (2004) 078},
\href{http://arxiv.org/abs/hep-ph/0402175}{{\ttfamily arXiv:hep-ph/0402175}}.

\bibitem{Schwetz:2011qt}
T.~Schwetz, M.~Tortola, and J.~Valle, ``{Global neutrino data and recent
  reactor fluxes: status of three-flavour oscillation parameters},''
  \href{http://dx.doi.org/10.1088/1367-2630/13/6/063004}{{\em New J.Phys.}
  {\bfseries 13} (2011) 063004},
  \href{http://arxiv.org/abs/1103.0734}{{\ttfamily arXiv:1103.0734 [hep-ph]}}.

\bibitem{prem}
A.~M. Dziewonski and D.~L. Anderson, ``{Preliminary reference earth model},''
  {\em Phys. Earth Planet. Interiors} {\bfseries 25} (1981) 297--356.

\bibitem{Donini:2003vz}
A.~Donini, D.~Meloni, and S.~Rigolin, ``{Clone flow analysis for a theory
  inspired neutrino experiment planning},''
  \href{http://dx.doi.org/10.1088/1126-6708/2004/06/011}{{\em JHEP} {\bfseries
  0406} (2004) 011},
\href{http://arxiv.org/abs/hep-ph/0312072}{{\ttfamily arXiv:hep-ph/0312072
  [hep-ph]}}.

\bibitem{Bishai:2012ss}
M.~Bishai, M.~Diwan, S.~Kettell, J.~Stewart, B.~Viren, {\em et al.},
  ``{Neutrino Oscillations in the Precision Era},''
\href{http://arxiv.org/abs/1203.4090}{{\ttfamily arXiv:1203.4090 [hep-ex]}}.

\bibitem{Huber:2009cw}
P.~Huber, M.~Lindner, T.~Schwetz, and W.~Winter, ``{First hint for CP violation
  in neutrino oscillations from upcoming superbeam and reactor experiments},''
  \href{http://dx.doi.org/10.1088/1126-6708/2009/11/044}{{\em JHEP} {\bfseries
  0911} (2009) 044}, \href{http://arxiv.org/abs/0907.1896}{{\ttfamily
  arXiv:0907.1896 [hep-ph]}}.

\bibitem{Ayres:2004js}
{\bfseries NOvA Collaboration} Collaboration, D.~Ayres {\em et al.}, ``{NOvA:
  Proposal to build a 30 kiloton off-axis detector to study nu(mu)
  $\rightarrow$ nu(e) oscillations in the NuMI beamline},''
  \href{http://arxiv.org/abs/hep-ex/0503053}{{\ttfamily arXiv:hep-ex/0503053
  [hep-ex]}}. Updated version of 2004 proposal. Higher resolution version
  available at Fermilab Library Server.

\bibitem{Yang2004}
{\bfseries NOvA} Collaboration, T.~Yang and S.~Woijcicki, ``Study of physics
  sensitivity of $\nu_mu$ disappearance in a totally active version of NoVA
  detector,'' \href{http://arxiv.org/abs/Off-Axis-Note-SIM-30}{{\ttfamily
  Off-Axis-Note-SIM-30}}.

\bibitem{Fechner2006}
M.~Fechner, {\em D\'etermination des performances attendues sur la recherche de
  l'oscillation $\nu_\mu\rightarrow\nu_e$ dans l'experi\'ence T2K depuis
  l'\'etude des donn\'ees recueilles dans l'\'experience K2K}.
\newblock PhD thesis, Universit\'e Paris VI, 2006.

\bibitem{Kato}
``Neutrino 2008.''. http://www.t2k.org/docs/talk/index\_html.

\bibitem{Huber:2002mx}
P.~Huber, M.~Lindner, and W.~Winter, ``{Superbeams versus neutrino
  factories},'' \href{http://dx.doi.org/10.1016/S0550-3213(02)00825-8}{{\em
  Nucl.Phys.} {\bfseries B645} (2002) 3--48},
  \href{http://arxiv.org/abs/hep-ph/0204352}{{\ttfamily arXiv:hep-ph/0204352
  [hep-ph]}}.

\bibitem{Ishitsuka:2005qi}
M.~Ishitsuka, T.~Kajita, H.~Minakata, and H.~Nunokawa, ``{Resolving neutrino
  mass hierarchy and CP degeneracy by two identical detectors with different
  baselines},'' \href{http://dx.doi.org/10.1103/PhysRevD.72.033003}{{\em
  Phys.Rev.} {\bfseries D72} (2005) 033003},
  \href{http://arxiv.org/abs/hep-ph/0504026}{{\ttfamily arXiv:hep-ph/0504026
  [hep-ph]}}.

\bibitem{mary}
M.~Bishai.
\newblock {Private communication}.

\bibitem{lisa}
L.~Whitehead.
\newblock {Private communication}.

\bibitem{Longhin:2010zz}
A.~Longhin, ``{Optimization of neutrino fluxes for European super-beams},''
{\em PoS} {\bfseries ICHEP2010} (2010) 325.

\bibitem{Agarwalla:2011hh}
S.~K. Agarwalla, T.~Li, and A.~Rubbia, ``{An Incremental approach to unravel
  the neutrino mass hierarchy and CP violation with a long-baseline Superbeam
  for large $\theta_{13}$},'' \href{http://arxiv.org/abs/1109.6526}{{\ttfamily
  arXiv:1109.6526 [hep-ph]}}.

\bibitem{LAGUNAmeeting}
F.~Di~Lodovico, ``Update WP5 LAr working group,''. Talk at the general LAGUNA
  meeting in Paris, 2012.

\bibitem{Laing:2010zza}
A.~B. Laing, ``{Optimisation of detectors for the golden channel at a neutrino
  factory},''. Ph.D. Thesis (Advisor: F. J. Paul Soler).

\end{thebibliography}

\providecommand{\href}[2]{#2}\begingroup\raggedright\endgroup

\end{document}